\documentclass[aps,10pt,prb,twocolumn,superscriptaddress,showpacs,floatfix]{revtex4-2}

\usepackage[usenames,dvipsnames]{xcolor}
\usepackage{amssymb}
\usepackage{graphicx}
\usepackage{amsmath}
\usepackage[bookmarks=true,colorlinks,citecolor=blue,urlcolor=blue]{hyperref}
\usepackage[utf8]{inputenc}
\usepackage[english]{babel}

\newcommand{\dd}{\mathrm{d}}
\newcommand{\CL}{\mathrm{cl}}

\begin{document}

\title{Many-body parametric resonances in the driven sine-Gordon model}

\newcommand{\TUM}{\affiliation{Department of Physics, Technical University of Munich, 85748 Garching, Germany}}
\newcommand{\MCQST}{\affiliation{Munich Center for Quantum Science and Technology (MCQST), Schellingstr. 4, 80799 M{\"u}nchen, Germany}}
\newcommand{\Harvard}{\affiliation{Department of Physics,  Harvard University,  Cambridge,  MA 02138,  USA}}
\newcommand{\Geneva}{\affiliation{DQMP,  University  of  Geneva,  24  Quai  Ernest-Ansermet,  CH-1211  Geneva,  Switzerland}}
\newcommand{\Salerno}{\affiliation{Dipartimento di Fisica ”E.R. Caianiello”, Universit\`{a} degli Studi di Salerno and CNR-SPIN c/o University of Salerno,Via  Giovanni  Paolo  II,  132,  I-84084  Fisciano  (Sa),  Italy}}
\newcommand{\INFN}{\affiliation{INFN,  Sezione  di  Napoli,  Gruppo  collegato  di  Salerno,  I-84084  Fisciano  (SA),  Italy}}
\newcommand{\Lyon}{\affiliation{ENSL, CNRS, Laboratoire de physique, F-69342 Lyon, France}}
\newcommand{\KITP}{\affiliation{Kavli Institute for Theoretical Physics, University of California, Santa Barbara, CA 93106, USA}}
\newcommand{\ETH}{\affiliation{Institute for Theoretical Physics, ETH Zurich, 8093 Zurich, Switzerland}}
\newcommand{\li}[1]{{\color{black} #1}}

 \author{Izabella Lovas} \TUM \MCQST \KITP
 \author{Roberta Citro} \Salerno \INFN
 \author{Eugene Demler} \Harvard \ETH
 \author{Thierry Giamarchi} \Geneva
 \author{Michael Knap} \TUM \MCQST
 \author{Edmond Orignac} \Lyon
 
\begin{abstract}
    We study a quantum many-body variant of the parametric oscillator, by investigating the driven sine-Gordon model with a modulated tunnel coupling via a semi-classical Truncated Wigner Approximation (TWA).  We first analyze the parametric resonant regime for driving protocols that retain our model gapped, and compare the TWA to a \li{Time-Dependent Gaussian Variational Ansatz (TGVA)}. We then turn to a drive which closes the gap, resulting in an enhanced energy absorption. While the TGVA approach breaks down in this regime, we can apply TWA to explore the dynamics of the mode-resolved energy density, and the higher-order  correlations between modes in the prethermal heating regime. For weak driving amplitude, we find an exponentially fast energy absorption in the main resonant mode, while the heating of all remaining modes is almost perfectly suppressed on short time scales. At later times, the highly excited main resonance provides effective resonant driving terms for its higher harmonics through the non-linearities in the Hamiltonian, and gives rise to an exponentially fast heating in these particular modes. We capture the strong correlations induced by these resonant processes by evaluating higher order connected correlation functions. Our results can be experimentally probed in ultracold atomic settings, with parallel one-dimensional quasi-condensates in the presence of a modulated tunnel coupling.
\end{abstract}

\maketitle

\section{Introduction}

In recent years, the advances in ultracold atomic experiments paved the way to study novel quantum many-body states, as well as non-equilibrium phenomena in unprecedented detail~\cite{coldatom_RevModPhys.80.885,coldatom_review2015,Gross995}. These systems provide an ideal platform to prepare interesting interacting many-body states in a controlled way, and allow for probing their dynamics. One of the paradigmatic many-body models investigated in these settings is the sine-Gordon model~\cite{sG}, realized by coupling two parallel quasi-one-dimensional bosonic condensates in a double well potential~\cite{Gritsev2007_PhysRevB.75.174511,Schweigler2017,PhaseRotation_PhysRevLett.120.173601,Schweigler2021, Kasper_PhysRevB.101.224102,foini_SG_coupledtubes,ruggiero_coupledtubes_inequivalents, Wybo2022}. The experimental realization of this model enabled the demonstration and characterization of non-Gaussian higher order correlations in the system, revealed the presence of topological soliton excitations through the full distribution of the spatially resolved relative phase between the two condensates~\cite{Schweigler2017}, and also lead to the observation of prethermalization in the non-equilibrium dynamics of the model~\cite{Schweigler2017,PhaseRotation_PhysRevLett.120.173601}. 

\begin{figure}
    \centering
    \includegraphics[width=\columnwidth]{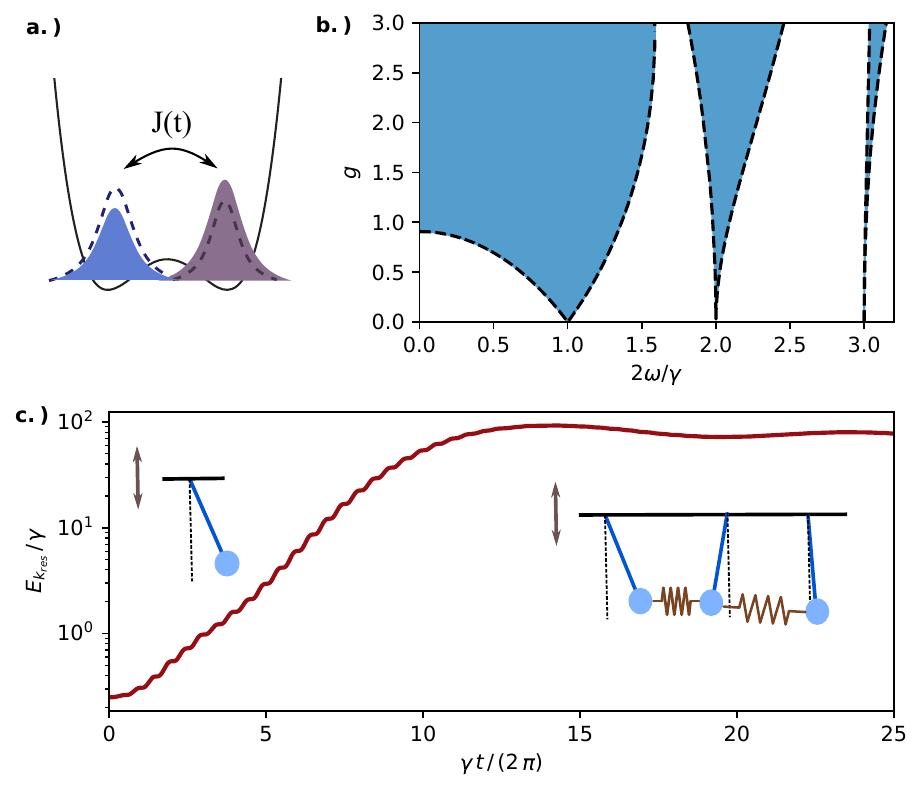}
    \caption{Parametric resonance in quasi-one-dimensional condensates in a modulated double well potential. \textbf{a.)} Experimental realization of the driven sine-Gordon model with coupled condensates in an oscillating trap. \textbf{b.)} Phase diagram of a classical parametric oscillator, revealing lobes of unstable heating regions (blue) as a function of frequency ratio $2\omega/\gamma$ and driving amplitude $g$, with natural frequency $\omega$ and driving frequency $\gamma$.  \textbf{c.)} Exponentially fast energy absorption in the resonant mode $k_{\rm res}$ of the sine-Gordon model. 
    Heating at short times is analogous to a single resonant parametric oscillator, whereas the saturation on longer time scales reflects a collection of oscillators, coupled through the interaction term.}
    \label{fig:setup}
\end{figure}

One of the interesting, only partially explored aspects of the out-of-equilibrium dynamics of the sine-Gordon model is the nature of transient states in the presence of a periodic drive. Previous works on this model  have focused either on regimes of small modulations, within the reach of linear response \cite{iucci_modulated_lattice,Citro2020_PhysRevResearch.2.033187} or on the limit of fast modulations, revealing a sharp crossover in the heating rate, separating regimes of strongly suppressed heating from regions with efficient energy absorption~\cite{CITRO2015694}. The less explored regime of slow driving frequencies offers many open questions regarding the quantum counterpart of well established classical phenomena, such as parametric resonance or amplification due to a resonant driving force~\cite{Bukov2015, Chandran2016, Weidinger2017, Herrmann_2017,Parametric_PhysRevX.7.021015,Parametric2D_PhysRevX.9.011047, Parametric_PhysRevX.10.011030,Fazzini_PhysRevLett.126.243401}, as well as about the universal aspects of the heating~\cite{Weidinger2017, Kuhlenkamp2020, Citro2020_PhysRevResearch.2.033187}. In particular,   the non-linear coupling between modes redistributes the energy absorbed by  the resonant mode, and can transiently stop the heating at a finite energy density, while at late times, when many-body scattering is taken into account, the system is expected to heat to an infinite temperature state~\cite{Weidinger2017}. This Floquet prethermalization effect has been demonstrated by investigating the total energy of various slowly driven many-body systems~\cite{Chandran2016, Weidinger2017}, however, the detailed mode-resolved energy absorption, shedding light on the dominant non-linearities in quantum systems, is much less explored. The driven sine-Gordon model provides an ideal, experimentally accessible platform to address these unresolved questions.

In this work, we investigate the mode-resolved energy absorption of the slowly driven sine-Gordon model in the presence of a modulated tunnel coupling, constituting a quantum many-body analogue of a parametric oscillator. This system is experimentally accessible in various cold atomic platforms, in particular,  by coupling two parallel, quasi-one-dimensional bosonic condensates through a modulated double well potential~\cite{Josephson1_PhysRevLett.95.010402,Levy2007,Berrada2013,Schweigler2017}, see Fig.~\ref{fig:setup}a. Alternatively, the modulated tunnel coupling can also be realized by introducing a Raman coupling between two internal states of the cold atoms~\cite{Zhang2013,HamnerSOC,OlsonSOC}.  In contrast to previous studies focusing on the limit of fast modulations~\cite{CITRO2015694}, we consider a slow driving frequency tuned to parametric resonance with one of the low energy modes of the static Hamiltonian. Hence, this system realizes a quantum many-body counterpart of the parametric oscillator, with non-linear coupling terms between modes. 

\li{A single classical parametric oscillator is described by the Hamiltonian function
\begin{equation}\label{eq:class}
    \mathcal{H}(t)=\dfrac{p^2}{2}+\dfrac{1}{2}\omega^2\left(1-2g\sin{(\gamma t)}\right)x^2,
\end{equation}
with natural frequency $\omega$, driving frequency $\gamma$, and driving amplitude $g$. The 
phase diagram of this classical  single oscillator is shown in Fig.~\ref{fig:setup}b}, displaying lobes of unstable heating regions as a function of the frequency $\gamma$ and amplitude $g$ of the drive~\cite{landau1976mechanics,arnold1988}.  In particular, a resonant drive leads to an exponentially fast energy absorption for arbitrarily weak driving amplitude. \li{The quantum many-body counterpart of this parametric oscillator can be obtained by first substituting $x^2/2\rightarrow 1-\cos(x)$ in Eq.~\eqref{eq:class} to get the Hamiltonian function of a driven pendulum, and then constructing and quantizing the Hamiltonian for a chain of driven pendulums, with the neighboring pendulums coupled through springs (see inset of Fig.~\ref{fig:setup}c). This procedure leads to the quantum sine-Gordon model (for an explicit expression as well as for an alternative derivation, starting from the Hamiltonian of Josephson coupled quasi-condensates, see Sec.~\ref{subsec:sGmodel}).
We demonstrate the exponential heating for the resonant mode in the full quantum many-body system, analogous to the heating of a single parametric oscillator,} in Fig.~\ref{fig:setup}c, evaluated with the semi-classical Truncated Wigner Approximation~\cite{POLKOVNIKOV20101790} (TWA), allowing for calculating the time evolution for various driving protocols. We also compare these results to another approach, the \li{Time-Dependent Gaussian Variational Ansatz (TGVA)}, whenever applicable. The TWA results show that the early stages of the dynamics can be well understood in terms of a \li{\textit{single} parametric oscillator, without considering its coupling to the other modes. However,  at longer time scales the non-linear coupling between modes becomes dominant and} cuts off the fast energy absorption, and leads to saturation, similarly to the prethermal behavior found in the driven O(N) model~\cite{Chandran2016, Weidinger2017}. Moreover, we find that the highly excited resonant mode can serve as an effective resonant drive for the higher harmonics through the non-linear coupling terms, leading to an efficient heating in these modes. This heating process occurs in separate stages. First, the main resonance is occupied, while the heating of other modes remains strongly suppressed. Second, the highly excited main resonance serves as a drive for its higher harmonics, with strength increasing in time as the resonant mode becomes even stronger populated. We show that these efficient coupling terms between the main resonance and its higher harmonics give rise to a characteristic pattern in higher-order correlation functions. 

The paper is organized as follows. We \li{briefly review} the derivation of the sine-Gordon model from the Hamiltonian of two coupled one-dimensional quasi-condensates in Sec.~\ref{subsec:sGmodel}, and  discuss the effect of a parametric drive within a linear approximation in Sec.~\ref{subsec:drive}. We first focus on driving protocols that keep the gap of the spectrum open in in Sec ~\ref{sec:gapped}, allowing us to compare TWA to the \li{TGVA}. We sketch the derivation of the \li{TGVA} in Sec.~\ref{subsec:scha}, and present the results of both approaches in Sec.~\ref{subsec:gappednumerics}. We turn to driving protocols which close the gap in Sec.~\ref{sec:numerics}, resulting in an enhanced energy absorption. In this regime we have to rely on TWA to follow the time evolution of the mode resolved energy absorption of the system. We concentrate on the heating of the main resonance and its higher harmonics in Sec.~\ref{subsec:harmonics}, and find deviations from the behavior of uncoupled oscillators. We construct a simple toy model to explain the main features our findings in Sec.~\ref{subsec:toymodel}, and test it by examining the correlations between modes in Sec.~\ref{subsec:correlations}.  We summarize our results and comment on the experimental realization in Sec.~\ref{sec:discuss}. Technical details are discussed in the appendices.

\section{Driven sine-Gordon model}\label{sec:system}

\subsection{Sine-Gordon description of coupled one-dimensional quasi-condensates}\label{subsec:sGmodel}

The dynamics of two  Josephson-coupled one-dimensional interacting quasi-condensates can be well approximated  by the sine-Gordon model. \li{In this subsection we briefly review this derivation from the literature. The reader only interested in the driving protocol considered in this paper may go directly to Sec.~\ref{subsec:drive}.} To this end, we consider two quasi-one-dimensional bosonic gases,  described by the Hamiltonian~\cite{Cazalilla_2004,sineG_PhysRevB.75.174511}
\begin{align}
H_{0}=&\sum_{j=1,2}\int d x\left\{ \frac{1}{2m}\partial_x\psi_{j}^{\dag}(x)\partial_x\psi_{j}(x)\right.\nonumber\\
& \left.\qquad+\frac{g}{2}\psi_{j}^{\dag}(x)\psi_{j}^{\dag}(x)\psi_{j}(x)\psi_{j}(x)\right\} ,\label{eq:LiebLiniger}
\end{align}
that are coupled by a time dependent Josephson tunneling term, 
\begin{equation}
H_{J}=-J(t)\int d x\left[\psi_{1}^{\dag}(x)\psi_{2}(x)+\psi_{2}^{\dag}(x)\psi_{1}(x)\right]\,.\label{HMicroscopic}
\end{equation}
Here $\psi_{1}(x),\psi_{2}(x)$ stand for the bosonic fields of the two
quasi-condensates, $g$ denotes the effective one dimensional
interaction, depending sensitively on the shape of the transverse trapping potential, $J(t)$ is the modulated tunneling amplitude, and we have set $\hbar=1$. For simplicity, we focus on homogeneous condensates, moreover, we use periodic boundary conditions in the numerical simulations. We will discuss the effects of open boundary conditions where relevant. 

It is convenient to rewrite the field operators in terms of the conjugate phase $\varphi_{j}(x)$ and density $\rho_{j}(x)$ \li{operators}
\begin{equation}
\psi_{j}(x)=\sqrt{\rho_{j}(x)}\;e^{i\varphi_{j}(x)}\;.\label{eq:phase_rep}
\end{equation}
Substituting Eq. \eqref{eq:phase_rep}
into the Hamiltonian, \li{and performing an expansion in terms of the small density fluctuations on the top of the homogeneous mean density of the quasi-condensates, as well as in terms of the fluctuations in the gradient of the phase, leads to  a hydrodynamical description of the system}~\cite{bosonization_PhysRevLett.47.1840,giamarchi2004quantum}. To leading order in the density and phase fluctuations,
the relative phase $\varphi=\varphi_{2}-\varphi_{1}$ and density  $\delta\rho=\rho_{2}-\rho_{1}$ decouples
from the total phase and density, $\varphi_{1}+\varphi_{2}$ and $\rho_{1}+\rho_{2}$. The dynamics of the relative coordinates is governed by the sine-Gordon Hamiltonian, see appendix \ref{app:bosonization} and Refs.~\onlinecite{Gritsev2007_PhysRevB.75.174511,DallaTorre2013_PhysRevLett.110.090404}, 
\begin{equation}\label{eq:H_cont}
H=\frac{c}{2}\int d x\left\{ \frac{\pi}{K}\delta\rho^{2}+\frac{K}{\pi}\left(\partial_{x}\varphi\right)^{2}\right\} -2J(t)\rho_{0}\int d x\cos\varphi\,,
\end{equation}
with $c$ denoting the sound velocity, $K$ the Luttinger parameter, and $\rho_{0}$ the average density of the homogeneous gas. Here, the Luttinger parameter $K$ characterizes the strength of interactions, with $K\rightarrow\infty$ and $K=1$ representing the non-interacting limit and the limit of hard-core bosons, respectively. In the absence of tunnel-coupling, $J(t)\equiv 0$, Hamiltonian ~\eqref{eq:H_cont} reduces to a Luttinger-liquid with a linear spectrum,
\begin{equation}\label{eq:ck}
 \varepsilon_k =c k.
\end{equation}
By contrast, a non-zero static tunnel-coupling, $J(t)\equiv J_0$, introduces a gap to the dispersion relation. For large $J_0$, where the phase is pinned around $\varphi(x)\sim 0$, the expansion of the cosine function gives the approximation
\begin{equation}\label{eq:ekgap}
    \varepsilon_k^{\rm gap} \approx \sqrt{c^2k^2+\Delta_0^2},
\end{equation}
with 
\begin{equation}\label{eq:egap}
  \Delta_0 = \sqrt{\dfrac{2\pi J_0\rho_0 c}{K}}.  
\end{equation}

In the numerical calculations, we focus on the lattice-regularized version of Eq. \eqref{eq:H_cont}, obtained by introducing a lattice spacing $a$, and particle number operators $n_j=a\,\delta\rho(ja)$,
\begin{align}
H_{\mathrm{Lat}}=&\frac{ c}{2}\sum_{j=1}^{N_{s}}\left(\frac{\pi}{Ka}n_{j}^{2}+\frac{K}{\pi a}\left(\varphi_{j}-\varphi_{j-1}\right)^{2}\right)\nonumber \\
&-2J(t)\rho_{0}a\sum_{j=1}^{N_{s}}\cos\varphi_{j}\;.\label{eq:HLat}
\end{align}
Here $N_s$ denotes the number of lattice sites, and the particle number and phase operators satisfy the canonical commutation relations
\begin{equation*}
    [n_i,\varphi_j]=-i\delta_{i,j}.
\end{equation*}
The model~(\ref{eq:HLat}) is known as the quantum Frenkel-Kontorova\cite{FK1938,hu_li_2000} chain. 

Our main tool used for exploring the dynamics of Hamiltonian~\eqref{eq:HLat} is the semi-classical truncated Wigner approximation (TWA). This approach allows us to calculate time-dependent expectation values and correlations by sampling the classical phase and particle number variables $\{\varphi_j,n_j\}$ randomly at $t=0$, according to the Wigner distribution of the initial state, and by calculating the time evolution from the mean-field equations of motion~\cite{POLKOVNIKOV20101790}. \li{We note that TWA can only be rigorously justified on short time scales based on a perturbative expansion. At intermediate times, the approximation becomes uncontrolled, and its range of validity can only be estimated by direct comparison to more controlled numerical methods or analytical approaches. For quantum quenches in the sine-Gordon model, such a benchmark has been performed in various works~\cite{Sinatra_2002,POLKOVNIKOV20101790,DallaTorre2013_PhysRevLett.110.090404,HorvathPhysRevA.100.013613}, establishing TWA as a reliable approach up to intermediate times and comparatively weak interactions, for the regime that is the main focus of the paper. While none of these works studied a Floquet drive, due to the difficulties in applying well-controlled approximations to highly excited states, generic considerations suggest that TWA performs even better in the presence of a high energy density, rendering the dynamics more classical. These arguments support the expectation that TWA correctly captures the transient parametric heating dynamics studied in this paper.}

\subsection{Parametric drive}\label{subsec:drive}

We consider the sine-Gordon Hamiltonian ~\eqref{eq:HLat} in the presence of a periodically modulated tunnel coupling,
\begin{equation}\label{eq:Jt}
    J(t)=J_0+J_1\sin(\gamma t).
\end{equation}
First we comment on driving protocols where $J(t)>0$ at all times, i.e. the instantaneous spectrum remains gapped throughout the time evolution, in Sec.~\ref{sec:gapped}. In this case we can compare the semi-classical TWA to another widely used approximation, the \li{TGVA}. Then in Sec.~\ref{sec:numerics} we turn to the main focus of the paper, the case with zero static component, $J_0=0$. Here, the gap closes and the potential energy $\sim J(t)\cos{\varphi}$ changes sign in each half period of the drive, giving rise to an enhanced energy absorption. For that case the \li{TGVA} breaks down because the gap closes, however, the TWA is still well suited for studying the mode resolved energy absorption. Note that for $J_0=0$ the static part of the Hamiltonian is a purely quadratic Luttinger-liquid, displaying the spectrum ~\eqref{eq:ck}, and allowing to resolve the dynamics according to the modes of the static Hamiltonian.

In order to study the quantum many-body counterpart of the classical parametric oscillator, we tune the driving frequency $\gamma$ to parametric resonance with one of the low energy modes,
\begin{equation}\label{eq:res}
    \gamma = 2\,\varepsilon_{k_{\rm res}}
\end{equation}
with $k_{\rm res}=2\pi n_{\rm res}/N_s$, $n_{\rm res}\in \mathbb{Z}$. To gain more insight into the dynamics, it is convenient to consider the classical equation of motion for the phase field $\varphi_j$,
\begin{equation}\label{eq:eom}
    \partial_t^2\varphi_j=\dfrac{ c^2}{a^2}\left(\varphi_{j+1}+\varphi_{j-1}-2\varphi_{j}\right)-\dfrac{2\pi J(t)\rho_0 c }{K}\sin\varphi_j,
\end{equation}
and apply the  linear expansion $\sin\varphi_j\approx \varphi_j$. For simplicity, here we set the static tunnel coupling to zero, $J_0=0$. By performing a Fourier transformation and introducing the dimensionless time $\tau=\varepsilon_{k_{\rm res}} t$ and wave number $\tilde{k}=k/k_{\rm res}$, we obtain the following Mathieu equation~\cite{landau1976mechanics,Broer2004a}
\begin{equation}\label{eq:mathieu}
    \left(\partial_\tau^2+\tilde{k}^2+2\tilde{g}\sin(2\tau)\right)\varphi_{\tilde{k}} = 0.
\end{equation}
Here, we have introduced the dimensionless driving amplitude~\footnote{\li{The Mathieu equation is usually written with an opposite sign in front of the term $2\tilde{g}\sin{2\tau}$. We are using a different sign convention in Eq.~\eqref{eq:mathieu} to keep the dimensionless driving amplitude $\tilde{g}$ appearing in our numerical simulations positive.}}
\begin{equation}\label{eq:g}
    \tilde{g}=\dfrac{4\pi J_1\rho_0 c}{\gamma^2 K}.
\end{equation}
The Mathieu equation ~\eqref{eq:mathieu} has been obtained by neglecting the coupling  between modes, therefore  describes a collection of independent parametric oscillators. \li{The solutions of this differential equation have been studied extensively, and we review some of their important properties below. The reader only interested in our numerical results may go directly to Sec.~\ref{sec:gapped}.

The Mathieu equation \eqref{eq:mathieu}} can be solved exactly in terms of Mathieu functions~\cite{landau1976mechanics,Broer2004a}, and the solution takes the form $\varphi_{\tilde{k}}=e^{ i\nu(\tilde{k},\tilde{g})}P(\tau)$, where the function $P(\tau)$ is periodic in $\tau$. The energy absorption of mode $\tilde{k}$ is thus determined by the so-called Mathieu exponent $\nu(\tilde{k},\tilde{g})$. A negative imaginary part ${\rm Im} \nu(\tilde{k},\tilde{g}) < 0$ gives rise to exponentially fast heating, whereas a real $\nu(\tilde{k},\tilde{g})$ corresponds to a stable, oscillating solution. Inspecting the imaginary part of $\nu(\tilde{k},\tilde{g})$ as a function of $\tilde{k}$ and $\tilde{g}$ gives rise to the single-oscillator phase diagram, Fig.~\ref{fig:setup}b, with the substitution $\omega\leftrightarrow\tilde{k}$.

The Mathieu exponent $\nu(\tilde{k},\tilde{g})$ depends sensitively on $\tilde{k}$, characterizing the ratio of the natural frequency of the oscillator and the driving frequency, as well as on the dimensionless driving amplitude $\tilde{g}$, as depicted in Fig.~\ref{fig:setup}b. In the regime of weak driving amplitude, $\tilde{g}\lesssim 1$, sharp unstable regions with strong heating, ${\rm Im}\nu(\tilde{k},\tilde{g})<0$, appear in the vicinity of the main resonance $\tilde{k}=1$, satisfying the resonance condition~\eqref{eq:res}, as well as around its higher harmonics $\tilde{k}\in\mathbb{Z}$. In particular, the unstable region around the main resonance $\tilde{k}=1$ extends to
\begin{equation*}
    1-\tilde{g}<\tilde{k}^2<1+\tilde{g}.
\end{equation*}
The higher-order resonances around $\tilde{k}=2,3,...$ are weaker with a smaller $|{\rm Im}\nu(\tilde{k},\tilde{g})|$, thus the heating of these modes occurs on longer time scales then the heating of the main resonance. In contrast, in the regime of strong driving, $\tilde{g}>1$, the resonances become very broad, with large unstable regions displaying an exponential energy absorption. We note that a similar phase diagram holds in the more general case, $J_0>0$, with unstable lobes appearing around the main resonance and its higher harmonics, $\varepsilon_k/\varepsilon_{k_{\rm res}}\in\mathbb{Z}$.

\section{Driving protocols retaining the gap}\label{sec:gapped}

In this section, we consider driving protocols that keep the gap open throughout the time evolution, $J(t)>0$ for all $t$. In this case we can compare TWA to the \li{TGVA}, which neglects the higher order correlations between modes and breaks down for driving protocols where the gap closes, i.e. $J(t)=0$ at certain times $t$. Therefore, we expect that TWA is better suited for examining the energy absorption of the system in the limit of strong modulation, $J_1>J_0$. Nonetheless, \li{TGVA can be used as a consistency check for our TWA results} in certain limits.

\subsection{\li{Time-dependent Gaussian variational Ansatz}}\label{subsec:scha}

We briefly sketch the main ingredients of the \li{time-dependent Gaussian variational ansatz (TGVA) from the literature, before turning to our numerical results in Sec.~\ref{subsec:gappednumerics}. Further details on the method} are relegated to the appendices, appendices \ref{app:fk-scha}--\ref{app:adiabatic}. 

We consider the time-dependent sine-Gordon Hamiltonian density of Eq.~(\ref{eq:H_cont}), 
\begin{equation}
\mathcal{H}[\varphi,\rho] = \frac{\pi c}{2 K}\delta\rho^2 + \frac{cK}{2\pi}(\partial_x \varphi)^2 - 2J(t) \rho_0\cos \varphi,
\end{equation} Here, $\varphi(x)$ and $\delta \rho(x)=-i\delta/\delta\varphi(x)$ satisfy the canonical commutation relation. Deep in the gapped phase, the ground state of the system can be well approximated by a Gaussian functional of the field $\varphi(x)$, localized near one of the minima of the cosine potential~\cite{coleman_equivalence,suzumura1979collective}. In time-dependent problems starting from a localized state, $\varphi$ remains localized as long as the gap is finite. The time evolution of the system can then be approximated by a time-dependent Gaussian wavefunction introduced by Cooper {\em et al.}~\cite{Cooper2003}:
\begin{equation}
\Psi_v[\varphi(x)] = \mathcal{A}\exp\left(-\int_{x,y}\,\varphi(x)\left[\frac {G^{-1}_{x,y}} 4 - i\Sigma_{x,y}\right]\varphi(y) \right).
\end{equation}
Here $G_{x,y}$ denotes the connected two-point correlator, 
$$G_{x,y}=\langle \varphi(x)\,\varphi(y) \rangle  - \varphi_\CL(x)\varphi_\CL(y),$$
where $\varphi_\CL(x)$ is the classical expectation value of the field, $\langle \varphi(x) \rangle = \varphi_\CL(x)$. The correlator $\Sigma_{x,y}$ corresponds to the conjugate function of $G_{x,y}$, and \li{the prefactor $\mathcal{A} \sim  ({\mathrm{det}\,G})^{-1/4}$ ensures that the wave function remains normalized  at all times},
\begin{equation}
\langle \Psi_v|\Psi_v \rangle = \int \mathcal{D}[\varphi]\Psi_v[\varphi]^*\,\Psi_v[\varphi] = 1.
\end{equation}

The functions $G_{x,y},\Sigma_{x,y}$ constitute variational parameters that can be determined from the Dirac variational principle. To this end, it is convenient to  define an effective classical Lagrangian density,
\begin{eqnarray}
&& \mathcal{L}_\CL[G_{x,y}, \Sigma_{x,y}, \varphi_\CL(x), p_\CL(x)] = \nonumber \\
&& \int \mathcal{D}[\varphi]\,\Psi_v[\varphi]\left(i\partial_t - \mathcal{H}[\varphi,\partial/\partial \varphi]\right)\Psi_v[\varphi].
\end{eqnarray}
The equations of motion are obtained by making $S_\CL \equiv \int \dd t\int \dd x\,\mathcal{L}_\CL$ stationary. For a translationally invariant system, it is convenient the perform a Fourier transformation, $G_{k} \equiv \int_{-\Lambda}^\Lambda \dd k\,G_{x-y}\,e^{-ik(x-y)}$, with $\Lambda$ denoting a UV cutoff. A similar expression holds for $\Sigma_{k}$.  After some algebra, the saddle point condition $\delta S_\CL = 0$ leads to the following equations of motion,
\begin{subequations}\label{eq:EOM}
\begin{align}
\dot{G}_k &= \frac{4\pi c}{K}\,G_k\Sigma_k,\\
\label{eq:sigdot}
\dot{\Sigma}_k &= \frac{\pi c}{8 K}\, G^{-2}_k - \frac{2\pi c}{K}\,\Sigma_k^2 - \frac{cK}{2\pi}k^2 - J(t)\,Z(t)\,\rho_0.
\end{align}
\end{subequations}
Here the factor
\begin{equation}\label{eq:Zt}
Z(t) = \exp\left(-\frac{1}{2}\int_{-\Lambda}^\Lambda \frac{\dd k}{2\pi} \, G_k\right)
\end{equation}
describes a renormalization of the tunnel coupling by the phase fluctuations. An equivalent Gaussian approximation for the dynamics can be obtained by replacing the time dependent variational principle arguments sketched above by an alternative~\cite{nieuwkerk_self-consistent_2019,van_nieuwkerk_low-energy_2020} self-consistent time dependent harmonic approximation (SCTDHA); see App.~\ref{app:fk-scha}. In that approach, the non-quadratic term $-2J(t) \rho_0 \cos \varphi$ is replaced with $-2 J(t) \rho_0 \langle \cos \varphi\rangle (1-\varphi^2/2)$, making the operator equations of motion linear. The expectation value $\langle \cos \varphi \rangle$ is then calculated self-consistently. We provide details on the equivalence of the two approaches in appendix~\ref{app:tdva-scha}.

\subsection{Numerical results for gapped driving protocols}\label{subsec:gappednumerics}

In this section we compare the \li{TGVA} to the semi-classical TWA for gapped driving protocols, $J_1<J_0$. We note that the \li{TGVA} breaks down once the modulated Hamiltonian crosses a gapless point. Indeed, the closing of the gap amplifies the quantum fluctuations $G_k$, leading to the suppression of the renormalization factor, $Z(t)\rightarrow 0$, and ensuring that the gap remains closed for the rest of the time evolution. This effect is an artefact of the harmonic approximation. In the gapped regime $J_1<J_0$, however, the quantum fluctuations can remain bounded, such that $Z(t)$ stays finite at all times, and $\varphi$ remains localized (see the appendix \ref{app:gapRenorm} for more details). 

\begin{figure}
    \centering
    \includegraphics[width=\columnwidth]{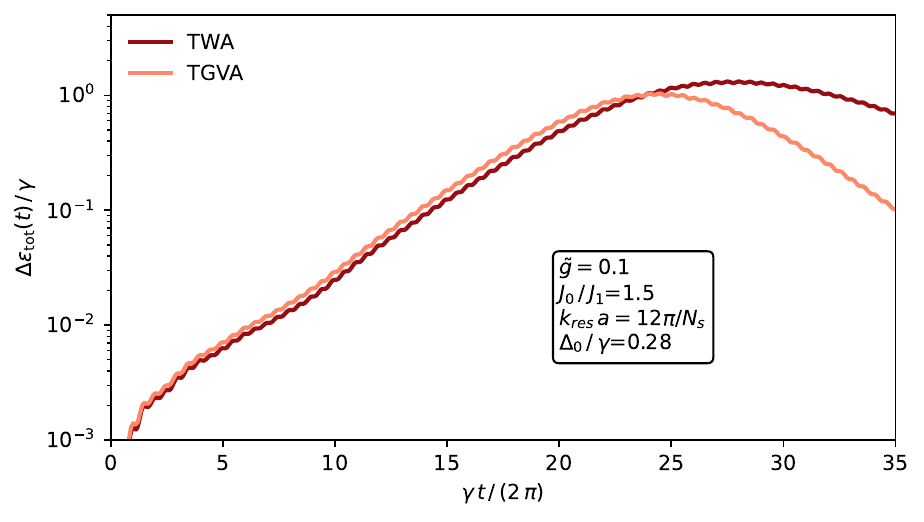}
    \caption{Parametric excitation of the sine-Gordon model in the gapped regime. We plot the absorbed total energy density $\Delta\varepsilon_{\rm tot}/\gamma$ of the static Hamiltonian on a log-linear scale as a function of rescaled time $\gamma t /(2\pi)$, obtained from the Truncated Wigner Approximation (TWA) and \li{the Time-Dependent Gaussian Variational Ansatz (TGVA)}. The energy absorption is  exponentially fast up to intermediate times, but eventually gets suppressed due to the coupling between modes. The two approaches show a good agreement in the regime of exponential heating, but at longer times \li{TGVA} predicts a stronger suppression in the energy absorption due to the overestimation of the coupling renormalization. We used $J_0/J_1=1.5$, $N_s=200$, $K=40$, $\tilde{g}=0.1$, $k_{\rm res}\,a=12\pi/N_s$, and $\Delta_0/\gamma=0.28$.}
    \label{fig:Etot}
\end{figure}

We consider the dynamics of the lattice Hamiltonian ~\eqref{eq:HLat}, starting from the approximate ground state of the system at $t=0$, obtained by expanding the cosine function up to second order, $\cos{\varphi_i}\approx 1-\varphi_i^2$. In the \li{TGVA}, this initial state corresponds to $\Sigma_k(t=0) = 0$ and
\begin{equation}\label{eq:tdva_green-eq}
G_k(t=0) = \frac{\pi}{2K}\,\frac{1}{\sqrt{4c^2/a^2\sin^2{(k a/2)} + \Delta_0^2}},
\end{equation}
with the approximate gap $\Delta_0$ given by Eq.~\eqref{eq:ekgap}. We focus on the total energy density of the static Hamiltonian,
\begin{equation*}
    \varepsilon_{\rm tot}(t)=\dfrac{\langle H_{\rm stat}\rangle (t)}{N_s},
\end{equation*}
where the static component $H_{\rm stat}$ is obtained by replacing $J(t)\rightarrow J_0$ in Eq.~\eqref{eq:HLat}. 

We set the driving frequency to satisfy the parametric resonance condition Eq.~\eqref{eq:res} for one of the low energy modes $k_{\rm res}$, and study the energy absorption $\Delta\varepsilon_{\rm tot}(t)=\varepsilon_{\rm tot}(t)-\varepsilon_{\rm tot}(0)$ using the \li{TGVA} and TWA approaches. The \li{TGVA} result is obtained by solving Eqs.~\eqref{eq:EOM} numerically, with initial conditions specified by Eq.~\eqref{eq:tdva_green-eq}. The Wigner distribution of the initial state, required for the TWA method, is also determined by Eq.~\eqref{eq:tdva_green-eq}, resulting in a Gaussian distribution for the initial phase and particle number fluctuations.

The results are shown in Fig.~\ref{fig:Etot} for a moderate modulation strength  $\tilde{g}=0.1$, tunnel coupling ratio $J_0/J_1=1.5$, and static gap $\Delta_0/\gamma=0.28$. The parametric resonance leads to an exponentially fast energy absorption at short and intermediate time scales, and we find a good agreement between the TWA and \li{TGVA} results in this regime. At longer times, however, the coupling between modes leads to the suppression of the energy absorption in the system, \li{with TGVA yielding a stronger suppression compared to TWA. This discrepancy stems from the rapid increase of phase fluctuations within TGVA, leading to strong phase decoherence and a pronounced renormalization of the tunnel coupling. A similar suppression of tunneling within a Gaussian approximation has been observed for quantum quenches in the sine-Gordon model~\footnote{I. Lovas, E. Demler, and G. Zar\'and, unpublished.}. In that scenario, it was shown that the Gaussian approach overestimates the renormalization of the tunnel coupling considerably, due to neglecting important mode coupling terms. In contrast, TWA yielded a moderate suppression of tunneling, in good agreement with more controlled numerical approaches. Based on this insight, we interpret the discrepancy in Fig.~\ref{fig:Etot} on longer time scales as the breakdown of TGVA, due to neglecting higher order correlations that are incorporated to the time evolution in TWA.}

\section{Driving protocols closing the gap}\label{sec:numerics}

In this section we turn to modulations in the regime $J_1>J_0$, resulting in the closing of the instantaneous gap during the time evolution. Here, we rely solely on TWA simulations, because the \li{TGVA} approach breaks down for these driving protocols. For simplicity, we focus on $J_0=0$, such that the static Hamiltonian is a quadratic Luttinger liquid, convenient for studying the mode resolved energy absorption of different modes $k$. First we discuss the heating in the main resonance, satisfying Eq.~\eqref{eq:res}, and its higher harmonics in Sec.~\ref{subsec:harmonics}. Then we present a simple toy model capturing the most relevant coupling terms in Sec.~\ref{subsec:toymodel}, and tests its predictions in Sec.~\ref{subsec:correlations}.

\subsection{Heating of main resonance and its higher harmonics}\label{subsec:harmonics}

The Mathieu equation~\eqref{eq:mathieu}, describing the dynamics of the sine-Gordon model up to linear order,  predicts sharp resonances for a parametric drive satisfying the resonance condition~\eqref{eq:res} around the main resonance and its higher harmonics, $\tilde{k}\in\mathbb{Z}$, for weak driving amplitude $\tilde{g}<1$. To test these predictions, and to identify the effect of non-linear mode couplings in the full many-body dynamics, we investigate the mode-resolved energy absorption of the sine-Gordon model, by applying TWA. As before, we focus on the lattice-regularized model, Eq.~\eqref{eq:HLat}, where the modulated tunnel coupling $J(t)$ is given by Eq.~\eqref{eq:Jt} with $J_0=0$. This Hamiltonian is quadratic at $t=0$, and we initialize the system in its ground state.

The quadratic part of the lattice Hamiltonian ~\eqref{eq:HLat} can be easily diagonalized by Fourier transformation, yielding the spectrum
\begin{equation}
\varepsilon_{k}=\frac{2 c}{a}\left|\sin\frac{ka}{2}\right|\,,\label{eq:LatticeDispRel}
\end{equation}
reducing to the linear Luttinger-liquid spectrum ~\eqref{eq:ck} for small wave numbers $k\ll1/a$. We tune the driving frequency to parametric resonance with one of the low-energy modes according to Eq.~\eqref{eq:res}, and consider the energy absorption of different modes,
\begin{equation}\label{eq:Ekt}
    E_k(t)= c\left(\frac{\pi}{2Ka}\langle n_{k}n_{-k}\rangle(t)+\frac{2K}{\pi a}\sin^2\frac{ka}{2}\langle\varphi_{k}\varphi_{-k}\rangle(t)\right).
\end{equation}
Here, we defined $n_k=1/\sqrt{N_s}\sum_j n_je^{-ijak}$, with a similar relation for $\varphi_k$. We can then apply the semi-classical truncated Wigner approximation~\cite{POLKOVNIKOV20101790}, for evaluating the mode-resolved energy, Eq.~\eqref{eq:Ekt}, as a function of time. Below we parametrize the modes by the dimensionless wave number introduced in Sec.~\ref{subsec:sGmodel}, $\tilde{k}\equiv k/k_{\rm res}$.

\begin{figure}
    \centering
    \includegraphics[width=\columnwidth]{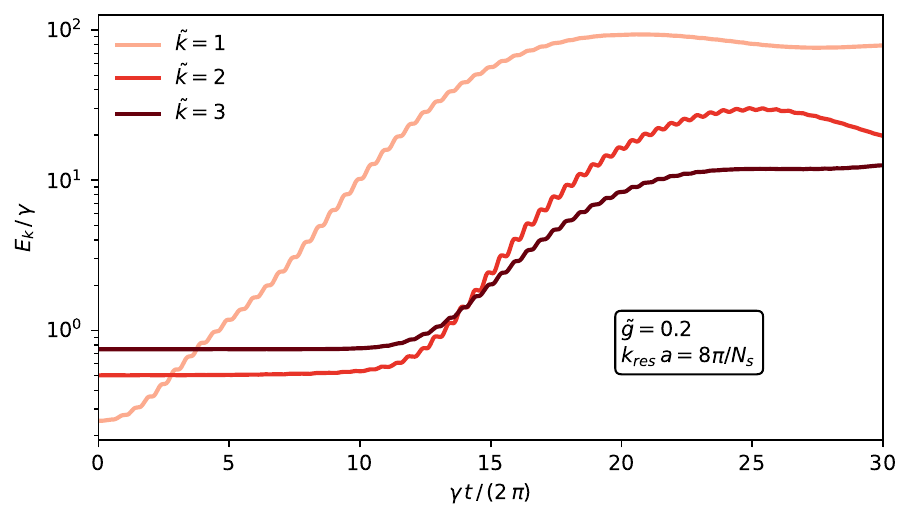}
    \caption{Parametric excitation of the resonant mode and its higher harmonics. We plot the energy $E_k$ on a log-linear scale as a function of rescaled time $\gamma t /(2\pi)$ for modes $\tilde{k}\equiv k/k_{\rm res}=1,2,3$. The main resonance $\tilde{k}=1$ shows an exponentially fast energy absorption due the external parametric drive. The energy absorption in higher harmonics is suppressed on short time scales, \li{but crosses over to an intermediate regime with efficient heating at later times, consistent with exponential increase, due to an effective drive provided by the non-linear coupling to the highly-excited main resonance.} We used $N_s=200$, $K=40$, $\tilde{g}=0.2$, and $k_{\rm res}\,a=8\pi/N_s$.}
    \label{fig:Ekt}
\end{figure}

We plot the energy Eq.~\eqref{eq:Ekt} for the main resonance $\tilde{k}=1$, as well as the higher harmonics $\tilde{k}=2$ and $\tilde{k}=3$ as a function of time in Fig. \ref{fig:Ekt} for a moderate modulation strength $\tilde{g}=0.2$. We find exponentially fast energy absorption in the main resonance $\tilde{k}=1$ already at early times, in accordance with the linear Mathieu equation discussed in Sec.~\ref{subsec:drive}. At later times, the energy saturates due to the strong coupling to other modes. However, the heating of the higher harmonics $\tilde{k}=2$ and $\tilde{k}=3$ differs substantially from the dynamics of uncoupled oscillators. In contrast to a higher order parametric resonance, displaying a steady, albeit slow, heating,  the energy absorption in modes $\tilde{k}=2$ and $\tilde{k}=3$ is almost completely suppressed at early times. \li{We then observe a crossover to an unstable regime at intermediate times, where the numerical results point towards an exponentially fast heating, followed by an eventual saturation at even later times.} This peculiar behavior is a consequence of the mode-coupling in the driven sine-Gordon Hamiltonian.

\subsection{Simplified model for mode-coupling}\label{subsec:toymodel}

In order to identify  the most relevant coupling terms between modes, and to shed more light to the results shown above, we present a simplified toy model for the dynamics, focusing on the main resonance and its higher harmonics. To this end, we consider Eq.~\eqref{eq:eom}, and apply the expansion $\sin\varphi_j\approx \varphi_j-\varphi_j^3/6$, thereby keeping the lowest order non-linearity in the equation of motion. By applying a Fourier transformation and changing to the dimensionless variables introduced in Sec.~\ref{subsec:drive}, we obtain
\begin{align}\label{eq:eom_nonlin}
    &\left(\partial_\tau^2+\tilde{k}^2+2\tilde{g}\sin(2 \tau)\right)\varphi_{\tilde{k}}=\nonumber\\
    &\qquad\qquad\dfrac{\tilde{g} }{6 N_s}\sin(2 \tau)\sum_{\tilde{k}_1,\tilde{k}_2}\varphi_{\tilde{k}_1}\varphi_{\tilde{k}_2}\varphi_{\tilde{k}-\tilde{k}_1-\tilde{k}_2}.
\end{align}
From our numerical results we conclude that the energy absorption at short and intermediate times is dominated by the main resonance and a few higher harmonics for weak drives. Therefore, we keep only the resonant indices $\tilde{k}\in\mathbb{Z}$ in Eq.~\eqref{eq:eom_nonlin} and drop the others. We  focus on the dynamics of \li{the main resonance $\tilde{k}=1$, and} the two lowest harmonics $\tilde{k}=2$ and $\tilde{k}=3$, and identify the most important coupling terms on the right hand side of Eq.~\eqref{eq:eom_nonlin}.

\li{Due to the rapid heating of the main resonance $\tilde{k}=1$, and because of the small population of modes $\tilde{k}=2$ and $\tilde{k}=3$ on short time scales, to leading order we can neglect all coupling terms to higher harmonics in the equation of motion for mode $\tilde{k}=1$. This reasoning leads to the following simplified equation for the main resonance,
\begin{equation}\label{eq:toy1}
   \left(\partial_\tau^2+1+2\tilde{g}\sin(2 \tau)\right)\varphi_1=\dfrac{\tilde{g}}{2N_s}\sin(2\tau)|\varphi_1|^2\varphi_1. 
\end{equation}
Turning to the second harmonics  $\tilde{k}=2$, we expect that the coupling to the main resonance $\tilde{k}=1$, the mode with the largest heating rate, will provide the dominant contribution to the energy absorption of mode $\tilde{k}=2$ on short times scales.} These considerations lead to the following equation of motion for mode $\tilde{k}=2$,
\begin{equation}\label{eq:eom_k2}
   \left(\partial_\tau^2+4+2\tilde{g}\sin(2 \tau)\right)\varphi_2=\dfrac{\tilde{g}}{N_s}\sin(2\tau)|\varphi_1|^2\varphi_2. 
\end{equation}
As noted in Sec.~\ref{subsec:drive}, the linear Mathieu equation for $\tilde{k}=2$ gives rise to a weak resonance for $\tilde{g}\lesssim 1$ (left hand side of Eq.~\eqref{eq:eom_k2}). The mode heats up exponentially as a function of time, however, the time scale associated with this heating is much longer than the typical time scale of energy absorption for the main resonance, $\tilde{k}=1$. Consequently, the contribution from this direct resonance is negligible at the early and intermediate stages of the dynamics. In contrast, we find that the non-linear coupling term on the right hand side gives rise to an effective  resonant parametric drive for mode $\tilde{k}=2$, and leads to a fast heating. Note that the time dependence of this driving term is determined by $\sin(2\tau)|\varphi_1|^2$. In our units, the natural frequency of mode $\tilde{k}=1$ is 1, thus the average $\langle|\varphi_1|^2\rangle$ contains an oscillating contribution $\sim\cos(2\tau)$. Combining this with the external modulation $\sin(2\tau)$, we arrive at an effective parametric drive $\sim\sin(4\tau)\varphi_2$. By comparing to the resonance condition, Eq.~\eqref{eq:res}, we find that this drive satisfies the parametric resonance condition, giving rise to a strong first order resonance, accompanied by an exponentially fast heating for mode $\tilde{k}=2$.  Importantly, the amplitude of this effective drive depends on the occupation of the main resonance through the amplitude $|\varphi_1|^2$, leading to a weaker effect at short time scales, but becoming dominant at intermediate times, once mode $\tilde{k}=1$ has sufficiently heated up.

\li{To qualitatively verify  the reasoning presented in the previous paragraph, we compare the numerical solution of the simplified equations ~\eqref{eq:toy1} and ~\eqref{eq:eom_k2} to the full TWA dynamics. The results are plotted in Fig.~\ref{fig:Ektoy} for a moderate coupling strength $\tilde{g}=0.2$. For the main resonance $\tilde{k}=1$, Eq.~\eqref{eq:toy1} yields a larger heating rate than the full TWA solution, because we neglect the coupling and energy transfer to other modes. For the higher harmonics $\tilde{k}=2$, Eq.~\eqref{eq:eom_k2} gives a decreasing energy at early times, whereas the energy of the mode remains approximately constant in the full TWA. At later times, the coupling to the main resonance becomes dominant, giving rise to a steady energy absorption, pointing towards a mild exponential heating. The heating rate, however, is considerably smaller than the rate obtained from the TWA simulations.

\begin{figure}[t!]
    \centering
    \includegraphics[width=\columnwidth]{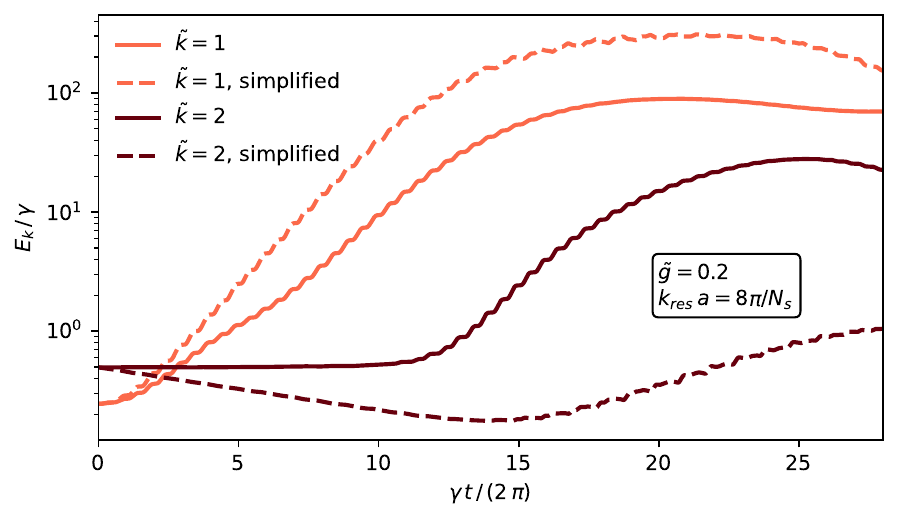}
    \caption{\li{Simplified description of heating. We compare the absorption predicted by the simplified equations ~\eqref{eq:toy1} and ~\eqref{eq:eom_k2} (dashed) to the full TWA time evolution (solid) for modes $\tilde{k}\equiv k/k_{\rm res}=1,2$. We plot the energy $E_k$ on a log-linear scale as a function of rescaled time $\gamma t /(2\pi)$. For the main resonance $\tilde{k}=1$, Eq.~\eqref{eq:toy1} predicts a faster heating than TWA, due to the absence of coupling to other modes. For the higher harmonics $\tilde{k}=2$, Eq.~\eqref{eq:eom_k2} yields an initial energy decrease. At later times, however, the emergent parametric drive provided by the main resonance gives rise to an efficient heating, pointing towards an exponentially fast energy absorption, albeit with a smaller rate than the one from the TWA solution. We used $N_s=200$, $K=40$, $\tilde{g}=0.2$, and $k_{\rm res}\,a=8\pi/N_s$.}}
    \label{fig:Ektoy}
\end{figure}

While the simplified equations ~\eqref{eq:toy1} and ~\eqref{eq:eom_k2} can not account for all details of the full TWA dynamics, they are able to capture some important qualitative features, and illustrate how the coupling to the main resonance induces fast, seemingly exponential heating in the second harmonics. To gain a similar qualitative understanding of the dynamics of higher harmonics, we now examine the analogous simplified equation of motion for the third harmonics $\tilde{k}=3$.}

Since lower harmonics are expected to display a larger heating rate for mode $\tilde{k}=3$, we identify the following dominant non-linear coupling terms, coupling $\tilde{k}=3$ to modes $\tilde{k}=1$ and $\tilde{k}=2$, 
\begin{align}\label{eq:eom_k3}
   &\left(\partial_\tau^2+9+2\tilde{g}\sin(2 \tau)\right)\varphi_3=\nonumber\\
   &\;\dfrac{\tilde{g}}{N_s}\sin(2\tau)\left(|\varphi_1|^2+|\varphi_2|^2\right)\varphi_3+\dfrac{\tilde{g}}{6 N_s}\sin(2\tau)\varphi_1^3. 
\end{align}
The first two terms on the right hand side give rise to effective parametric drives, whereas the last term acts as an external driving force for mode $\tilde{k}=3$. The first effective parametric drive, $\sim\sin(2\tau)|\varphi_1|^2$, becomes stronger rapidly due to the exponentially fast heating of mode $1$, and starts to induce an efficient energy absorption in mode $\tilde{k}=3$ at intermediate time scales. The second  parametric drive, $\sim\sin(2\tau)|\varphi_2|^2$, contains terms oscillating as $\sin(2\tau)\cos(4\tau)\sim\sin(6\tau)$, satisfying the resonance condition ~\eqref{eq:res} for mode $\tilde{k}=3$, and leading to a first order parametric resonance. Similarly to the effective resonant drive identified for mode $\tilde{k}=2$, the amplitude of the drive depends on the heating of a lower-lying mode. Therefore, the heating rate induced by this coupling becomes larger at intermediate time scales, when the occupation of the lower mode is sufficiently large. Finally, the third term on the right hand side of Eq.~\eqref{eq:eom_k3} amounts to an external diving force. Here $\varphi_1^3$ gives rise to oscillating terms of the form $e^{\pm 3 i \tau}$ and $e^{\pm i \tau}$, leading to an effective resonant drive $\sin(2\tau)e^{\pm i \tau}\sim e^{\pm 3 i \tau}$ for mode $\tilde{k}=3$. While a resonant external force with constant amplitude gives rise to a heating that is linear in time, the force provided by mode $\tilde{k}=1$ increases exponentially with time, as mode $\tilde{k}=1$ becomes more populated. Therefore, this term \li{may contribute towards an} exponentially fast heating of mode $\tilde{k}=3$ at intermediate time scales, similarly to the effective parametric drives discussed above.

Our simplified model provides a qualitative explanation for the results plotted in Fig.~\ref{fig:Ekt}. At short times, the main resonance starts to heat up exponentially fast due to the external parametric drive. In contrast, the direct energy absorption of the higher harmonics $\tilde{k}>1$ from the external drive is much less efficient, moreover, the non-linear coupling to mode $\tilde{k}=1$ is still weak due to the low population of this mode. Therefore, the heating of modes $\tilde{k}>1$ remains suppressed on short time scales. At intermediate time scales, the coupling between mode $\tilde{k}=1$ and its higher harmonics becomes dominant due to the high population of the main resonance. The strong effective parametric drive and  external driving force emerging from this coupling give rise to \li{a rapid heating in modes $\tilde{k}=2$ and $\tilde{k}=3$, consistent with an exponentially fast energy absorption. We note that the simplified equations, Eqs.~\eqref{eq:eom_k2} and ~\eqref{eq:eom_k3}, do not provide a rigorous evidence for exponential heating, since the population of the main resonance, and consequently the amplitude of the effective parametric drive, keeps changing on the time scales considered here. However, both the hand waving arguments presented above, and the numerical results obtained for the simplified model and for the full TWA dynamics, strongly suggest such an exponential heating regime.} Finally, at even later times the population of all of these modes saturates due to the coupling to the bath formed by the remaining modes.
 
Even though the simplified model yields a qualitative explanation for the results plotted in Fig.~\ref{fig:Ekt}, it does not capture the almost perfect suppression of heating in modes $\tilde{k}=2$ and $\tilde{k}=3$ at short time scales, instead predicting a slow decrease in energy \li{(see Fig.~\ref{fig:Ektoy})}.
This deviation between the simplified description and  the full quantum model is a direct consequence of the difference between the expansion $\sin\varphi\approx \varphi-\varphi^3/6$, and the exact function $\sin\varphi$ appearing in the mean field equation of motion, pointing towards the importance of the precise form of the many-body potential in the dynamics.

\subsection{Distribution of absorbed energy and higher-order correlations}\label{subsec:correlations}

\begin{figure}
    \centering
    \includegraphics[width=\columnwidth]{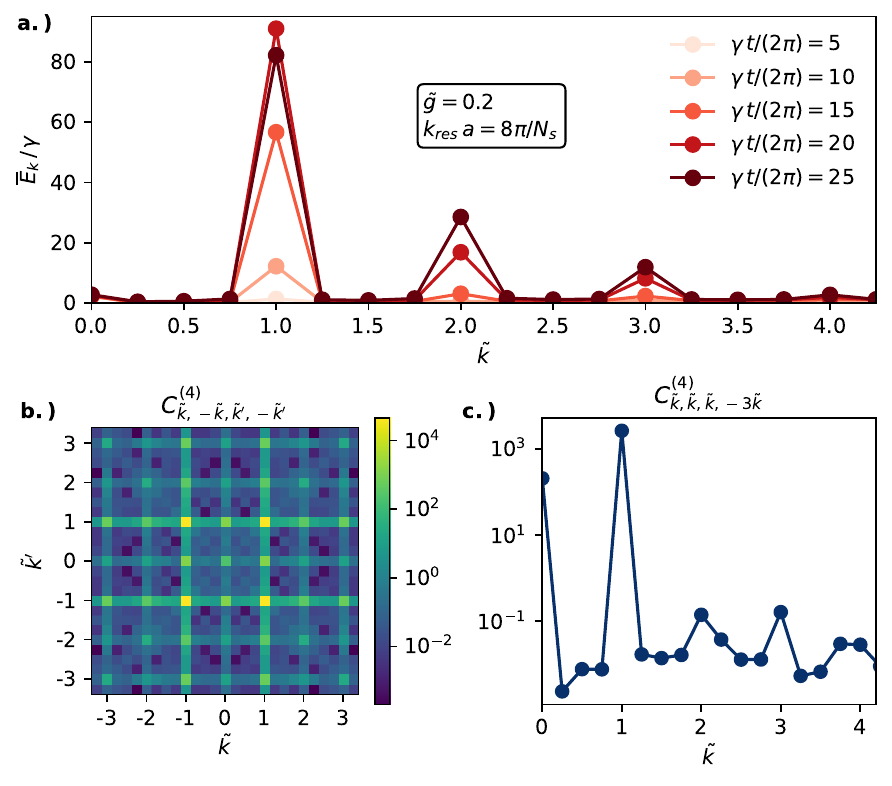}
    \caption{Energy profile and correlations for a weak drive. \textbf{a.)} Snapshots of the energy profile $\overline{E}_k$, averaged over one driving period, for different times, displaying sharp peaks at the main resonance and its higher harmonics. Exponential heating in the main resonance starts immediately, whereas the absorption in higher-order resonances is strongly suppressed at short times. At later times sharp peaks also emerge at the higher harmonics, as they absorb energy from the highly-excited main resonance.  \textbf{b.)} Higher-order correlations between modes obtained from the connected correlator $C^{(4)}_{k,-k,k^\prime,-k^\prime}$ as a function of dimensionless wave numbers $\tilde{k}=k/k_{\rm res}$ and  $\tilde{k}^\prime=k^\prime/k_{\rm res}$. \textbf{c.)} Higher-order correlators $C^{(4)}_{k,k,k,-3k}$ as a function of $\tilde{k}$, averaged over one driving period at time $\gamma t/(2\pi)=12.5$. The correlator $C^{(4)}_{k,-k,k^\prime,-k^\prime}$ displays strong correlations between the main resonance and the higher harmonics, $\tilde{k}\in\mathbb{Z},\tilde{k}^\prime\in\mathbb{Z}$, while $C^{(4)}_{k,k,k,-3k}$ reveals a sharp correlation peak between modes 1 and -3. We used $N_s=200$, $K=40$, $\tilde{g}=0.2$, and $k_{\rm res}\,a=8\pi/N_s$.}
    \label{fig:weak_drive}
\end{figure}

We examine the heating of different modes, as well as the higher order correlations in the system for weak driving amplitude in Fig~\ref{fig:weak_drive}. This allows us to test the predictions of the simplified model presented in the previous section in more detail. We will briefly comment on the case of stronger modulations at the end of the section.

To cancel the rapid oscillations stemming from the external drive, we average the energy over one period of the drive, 
\begin{equation}\label{eq:Ekt_av}
    \overline{E}_k(t)=\dfrac{1}{T}\int_{t-T/2}^{t+T/2}{\rm d}t^\prime E_k(t^\prime),
\end{equation}
with $T=2\pi/\gamma$. We plot snapshots of the mode-dependent energy $\overline{E}_k$ at different times for a weak driving amplitude $\tilde{g}=0.4$ in Fig~\ref{fig:weak_drive}a. As anticipated in Sec.~\ref{subsec:toymodel}, the heating at short and intermediate time scales is dominated by the main resonance and its higher harmonics, leading to  sharp peaks in the energy at wave numbers $\tilde{k}\in\mathbb{Z}$.  In accordance with the delay of the heating in higher harmonics discussed above, the resonance peaks around $\tilde{k}=2$ and $\tilde{k}=3$ remain strongly suppressed on short time scales, and become more pronounced at later times.

To further investigate the scope of the effective description discussed in Sec.~\ref{subsec:toymodel}, we also examine the higher-order density correlations in the system. \li{We note that for a Gaussian state, all connected correlation functions above second order vanish due to Wick's theorem. Therefore, by considering higher order correlators, we measure the deviation of the many-body wave function from a Gaussian state, and reveal a non-trivial correlated structure of the system. The lowest order non-Gaussian correlations are captured by fourth order correlators. Therefore, we} consider two different fourth-order correlation functions,
\begin{equation}\label{eq:C1}
    C^{(4)}_{k,-k,k^\prime,-k^\prime} = \langle n_{k}\, n_{-k}\,n_{k^\prime}\,n_{-k^\prime}\rangle_c,
\end{equation}
and
\begin{equation}\label{eq:C2}
    C^{(4)}_{k,k,k,k^\prime} = \langle n_{k}^3\, n_{k^\prime}\rangle_c,
\end{equation}
with $\langle ...\rangle_c$ standing for a connected correlator, and $n_k$ denoting the Fourier transform of the particle number operator, $n_k=1/\sqrt{N_s}\sum_j n_je^{-ijak}$. ~\li{The choice to focus on correlations between wave numbers $\{k, -k, k^\prime,-k^\prime\}$ and $\{k, k, k,k^\prime\}$ was motivated by our simplified model, Eqs.~\eqref{eq:eom_k2} and ~\eqref{eq:eom_k3}, implying potential strong correlations between these particular sets of wave numbers.} Due to translation invariance, $C^{(4)}_{k,k,k,k^\prime}$ is non-zero only for $k^\prime=-3k$, whereas  $C^{(4)}_{k,-k,k^\prime,-k^\prime}$ can be finite for arbitrary wave numbers $k$ and $k^\prime$. Similarly to Eq.~\eqref{eq:Ekt_av}, we average these correlators over one period of the drive.  

We show snapshots of the averaged correlators $C^{(4)}_{k,-k,k^\prime,-k^\prime}$ and $C^{(4)}_{k,k,k,-3k}$ at an intermediate time $\gamma t/(2/\pi)=12.5$, in the regime of weak modulation, $\tilde{g}\lesssim 1$, in Fig~\ref{fig:weak_drive}b and c. We find that $C^{(4)}_{k,-k,k^\prime,-k^\prime}$ displays sharp correlation peaks between the multiples of the resonant mode, $\tilde{k}\in\mathbb{Z},\tilde{k}^\prime\in\mathbb{Z}$. This behavior can be understood based on the simplified equations~\eqref{eq:eom_k2} and ~\eqref{eq:eom_k3}. The effective parametric drive appearing on the right hand side of these equations induces strong correlations of the type $C^{(4)}_{k,-k,k^\prime,-k^\prime}$ between the main resonance and the higher harmonics, $\tilde{k},\tilde{k}^\prime\in\mathbb{Z}$, in accordance with the results plotted in  Fig~\ref{fig:weak_drive}b. By contrast, the correlator $C^{(4)}_{k,k,k,-3k}$ reveals strong correlations between $n_{-3\tilde{k}}$ and $n_{\tilde{k}}^3$ for the main resonance $\tilde{k}=1$, as well as much weaker correlation peaks for the higher harmonics $\tilde{k}=2$ and $\tilde{k}=3$ (see Fig~\ref{fig:weak_drive}c). The sharp correlation peak at $\tilde{k}=1$ is consistent with the effective driving force appearing on the right hand side of Eq.~\eqref{eq:eom_k3}, providing a strong coupling between the operators $\varphi_1^3$ and $\varphi_{-3}$.

\begin{figure}
    \centering
    \includegraphics[width=\columnwidth]{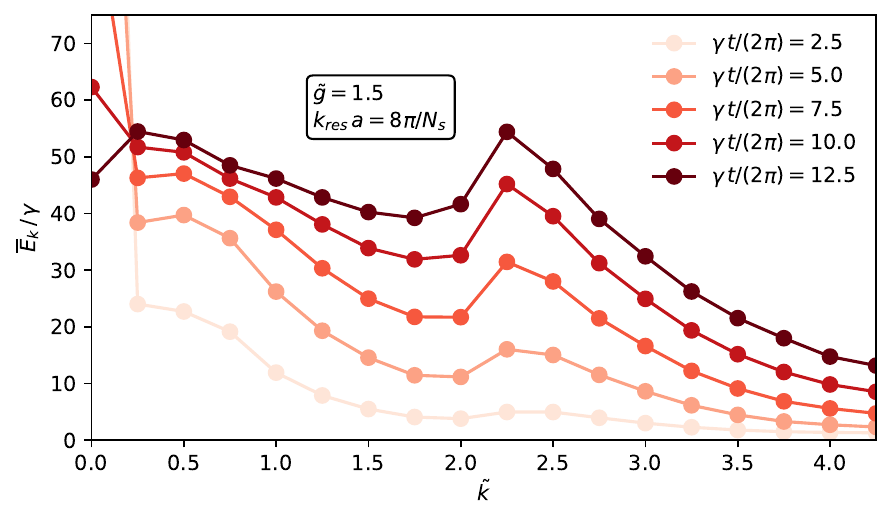}
    \caption{Energy profile for strong a drive.  Snapshots of the energy profile $\overline{E}_k$ for different times, averaged over one driving period, for a large modulation $\tilde{g}=1.5$. The broadening of the parametric resonance and the strong coupling between modes leads to a smoother energy profile than the one obtained for weak drives $\tilde{g}\lesssim 1$. The threshold $\tilde{k}_{\rm thres}\approx 2.25$ separates a broad region with fast heating from the more stable modes with slower energy absorption. We used $N_s=200$, $K=40$, and $k_{\rm res}\,a=8\pi/N_s$.}
    \label{fig:strong_drive}
\end{figure}

\begin{figure*}
    \centering
    \includegraphics[width=0.95\textwidth]{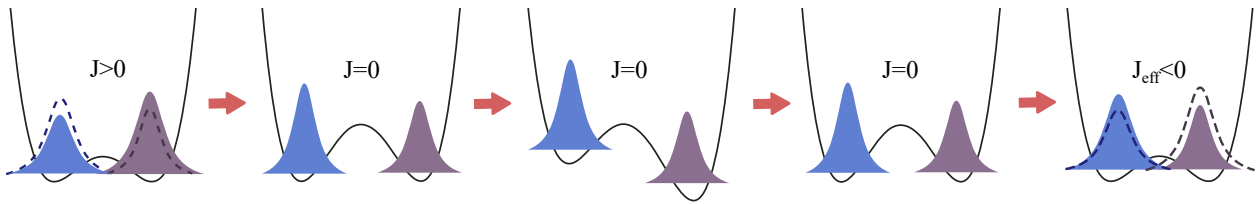}
    \caption{Proposed experimental protocol. The driven sine-Gordon model is experimentally accessible with quasi-one-dimensional condensates in a double-well potential, coupled through a modulated Josephson tunneling term. A tunnel-coupling oscillating around zero mean can be realized as follows: In the first half-period, the tunnel coupling is modulated as $J(t)=J_1\sin(\gamma t)$, $0\leq t\leq \pi/\gamma$. Upon reaching $J(t)=0$, the sign of the tunneling is reversed by implementing a global phase rotation $\pi$, through switching on a left-right asymmetry in the trapping potential. Once the desired phase difference $\pi$ has been imprinted, the asymmetry is switched off. Repeating after this rotation the same half-period $J(t)=J_1\sin(\gamma t)$, $0\leq t\leq \pi/\gamma$,  results in a an effective tunnel coupling $J_{\rm eff}(t)=-J_1\sin(\gamma t)$, completing the driving cycle.}
    \label{fig:protocol}
\end{figure*}

In the presence of a large driving amplitude $\tilde{g}> 1$, the heating dynamics is modified. We plot the mode-resolved energy profile $\overline{E}_k(t)$ for different times in Fig.~\ref{fig:strong_drive}, using the driving strength $\tilde{g}=1.5$.
In contrast to the case of a weak drive discussed above, a strong modulation leads to a broad parametric resonance, as well as to strong non-linear couplings between modes, resulting in a smoother energy profile $E_k$ at all times, without well defined resonance peaks. Instead, we observe a threshold at $\tilde{k}_{\rm thres}\approx 2.25$, above which the energy starts to fall off as a function of $\tilde{k}$. We can gain a qualitative understanding of this behavior based on the uncoupled Mathieu equations ~\eqref{eq:mathieu}, predicting a broadened first and second resonance in close proximity.
The mode-coupling modifies the boundary of these unstable lobes, merging them into a single broad heating region ${\rm Im}\,\nu(\tilde{k},\tilde{g})<0$ with exponentially fast heating below the threshold  $\tilde{k}_{\rm thres}$. While this perturbative reasoning explains the main features in Fig.~\ref{fig:strong_drive}, the details of the dynamics depend strongly on the pronounced coupling between the modes, and can only be obtained from a simulation accounting for the full many-body potential. 

For the strong drive considered in Fig.~\ref{fig:strong_drive}, the homogeneous mode $k=0$ heats up exponentially fast~\footnote{\li{The typical time scale of the exponential heating in mode $k=0$ is shorter than the time scale associated with the heating in modes $k>0$. The snapshots plotted in Fig.~\ref{fig:strong_drive} were chosen to reflect the change of population in modes $k>0$; at these time scales, the exponential heating of mode $k=0$ has already stopped, instead, the excitation is transferred gradually from $k=0$ to the higher modes $k>0$.}}, with a large heating rate, giving rise to a pronounced peak on short time scales (compare to the phase diagram of a single oscillator, Fig.~\ref{fig:setup}). At later times, this high excitation is transferred to the remaining modes, as can be observed in Fig.~\ref{fig:strong_drive}.

\li{\section{Outlook and  implications for experiments}}\label{sec:discuss}



We have considered a quantum many-body analogue of a classical parametric oscillator,  the sine-Gordon model in the presence of a modulated cosine potential. For weak driving amplitudes, we have found an exponentially fast heating of the main resonance, whereas the energy absorption of the higher-order resonances is suppressed on short time scales. On longer time scales the non-linear coupling terms in the Hamiltonian excite higher-oder resonances by effective parametric drives as well as external driving forces generated from the highly excited main resonance. Such mode coupling allow the higher resonances to absorb energy from the highly excited main resonance and gives rise to further sharp resonance peaks in the mode-resolved energy density. We have identified the most relevant couplings by exploring the higher-order correlations between modes, and constructed a simplified model to explain our findings. While this model provides a good qualitative understanding for our main results, it fails to describe some aspects of the dynamics, such as the almost perfect suppression of energy absorption at short times in all modes except the main resonance. This indicates the relevance of the full many-body potential in the time evolution.

\li{The modulated tunnel coupling can be experimentally realized in various ultracold atomic settings, with one possible platform relying on coupling two hyperfine states of the cold atoms by a Raman coupling~\cite{Zhang2013,HamnerSOC,OlsonSOC}. Here we discuss another possible realization of the driven sine-Gordon model, in a setting consisting of 
two parallel quasi-one-dimensional condensates in the presence of a modulated tunnel coupling.}  The one-dimensional description of this system, Eq.~\eqref{eq:LiebLiniger}, fails when the transverse modes of the double well potential cannot be neglected anymore. At late times, these transverse modes may become populated due to the strong heating, leading to deviations from the sine-Gordon description. For the parameters used in this work, the total energy density of the system remains small compared to the chemical potential $gN$, with $N$ denoting the total particle number (despite some modes being exponentially populated). Under these conditions, we expect that the transverse modes of the trap can be neglected on the time scales we consider. 

While realizing an oscillating coupling of the form of Eq. \eqref{eq:Jt} with a deformed double well potential is relatively straightforward in the regime $J_1<J_0$, reversing the sign of $J(t)$ is substantially more challenging. We propose the procedure schematically depicted in Fig.~\ref{fig:protocol} to realize the tunnel coupling $J(t)=J_1\sin(\gamma t)$.  The first half-period, $0\leq t\leq \pi/\gamma$, can be realized by deforming the double-well trapping potential. An effective sign change can then be implemented by changing the global phase difference,  $\varphi_1-\varphi_2$, of the now uncoupled condensates by $\pi$. In the absence of tunnel coupling, $J(t)=0$, the phase difference $\pi$ can be imprinted by switching on an energy difference $\epsilon$ between the two condensates, by adding a small left-right asymmetry to the double well trapping potential~\cite{Berrada2013,PhaseRotation_PhysRevLett.120.173601}. The energy difference results in a global phase accumulation $\Delta(\varphi_1-\varphi_2)=\epsilon t$, reaching the desired value $\pi$ at time $t=\pi/\epsilon$. Once the phase $\pi$ has been imprinted, the energy difference $\epsilon$ can be switched off. Repeating the same modulation sequence, $J(t)=J_1\sin(\gamma t)$, $0\leq t\leq \pi/\gamma$, leads to a time-dependent cosine potential with an opposite sign, $-J(t)\cos(\varphi_j + \pi) = J(t)\cos\varphi_j$, completing the first driving cycle. 

We note that we have performed our simulations with periodic boundary conditions, amounting to a homogeneous average density and phase, $\langle n_k\rangle=\langle\varphi_k\rangle=0$. \li{An experimental realization typically corresponds to open boundary conditions. The resulting boundary terms are relevant perturbations modifying the field theoretical description, and lead to Friedel oscillations superimposed on the modulations caused by the parametric drive~\cite{Citro_PhysRevResearch.2.033187}. Nevertheless, in a box potential the parametric instability could still be detected through the time dependent density and phase correlation functions in the bulk, $\langle \rho(x) \rho(x^\prime)\rangle(t)$ and $\langle \varphi(x) \varphi(x^\prime)\rangle(t)$, where the distance between $x,x^\prime$ and the edge of the system is much larger than the healing length, as well as through the higher-order correlators. Alternatively, boundary effects can be avoided by implementing a ring geometry.}

Besides the  experimental relevance of our results, studying the mode-resolved energy absorption of other slowly driven quantum many-body systems remains of interest. The distribution of the absorbed energy, and the emerging correlation patterns can shed light on the dominant coupling terms between quasi-particles, as well as on the role of conservation laws or kinetic constraints.


\textbf{Acknowledgements}. 
We thank J\"org Schmiedmayer for insightful discussions. We acknowledge support from the Deutsche Forschungsgemeinschaft (DFG, German Research Foundation) under Germany’s Excellence Strategy--EXC--2111--390814868, TRR80 and DFG grants No. KN1254/1-2 and No. KN1254/2-1, the European Research Council (ERC) under the European Union’s Horizon 2020 research and innovation programme (grant agreements No. 851161 and 771537), as well as the Munich Quantum Valley, which is supported by the Bavarian state government with funds from the Hightech Agenda Bayern Plus. This work was supported by the Gordon and Betty Moore Foundation through Grant GBMF8690 to UCSB, by the National Science Foundation under Grant No. NSF PHY-1748958, by the Harvard-MIT CUA,  AFOSR-MURI award FA95501610323, the ARO grant “Control of Many-Body States Using Strong Coherent Light-Matter Coupling in Terahertz Cavities”, and by the Swiss National Science Foundation under Division II.

\appendix

\section{Derivation of the bosonized Hamiltonian}\label{app:bosonization}

Using the notations of Ref.~\onlinecite{giamarchi2004quantum}, the low energy excitations of the decoupled Hamiltonian~(\ref{eq:LiebLiniger}) are described by the bosonized Hamiltonian
\begin{eqnarray*}
  \label{eq:boso-decoupl}
    H_0=\sum_{j=1,2} \int \frac{dx}{2\pi} \left[ u_* K_* (\pi \Pi_j)^2 + \frac {u_*} {K_*} (\partial_x \phi_j)^2\right],
\end{eqnarray*}
where $u_*$ is the velocity of excitations, $K_*$ the Luttinger parameters, $[\phi_j(x), \Pi_k(x')]=i\delta_{jk} \delta(x-x')$. As a result of Galilean invariance\cite{bosonization_PhysRevLett.47.1840},
\begin{equation*}
  u_* K_* =\frac{\pi \rho_0}{m}, 
\end{equation*}
while $K_*/u_*$ can be obtained from the compressibility of the Lieb-Liniger model. For weak coupling, we have the approximations
\begin{eqnarray*}
  u_*\simeq\sqrt{\frac{\rho_0 g}{m}}, \\
  K_*\simeq\pi \sqrt{\frac{\rho_0}{mg}}.  
\end{eqnarray*}
The time dependent Josephson tunneling term~(\ref{HMicroscopic}) has the bosonized expression
\begin{eqnarray*}
H_J= - 2J(t) \rho_0 \int dx \cos (\theta_1-\theta_2)(x),
\end{eqnarray*}
with $\partial_x \theta_j = \pi  \Pi_j$.
It is convenient to introduce symmetric and antisymmetric combinations
\begin{eqnarray*}
  \label{eq:sym-asym-def}
  \phi_r=\frac 1 {\sqrt{2}} (\phi_1 + r \phi_2), 
  \theta_r=\frac 1 {\sqrt{2}} (\theta_1 + r \theta_2)
\end{eqnarray*}
with $r=\pm$, and rewrite $H=H_++H_-$ with
\begin{eqnarray*}
  \label{eq:hplus}
&&  H_*=\int \frac{dx}{2\pi} \left[ u_* K_* (\pi \Pi_+)^2 + \frac {u_*} {K_*} (\partial_x \phi_+)^2\right],\\ 
  \label{eq:hminus}
  &&  H_-=\int \frac{dx}{2\pi} \left[ u_* K_* (\pi\Pi_-)^2 + \frac {u_*} {K_*} (\partial_x \phi_-)^2\right] \nonumber \\
  && - 2 J(t) \rho_0 \int \cos \sqrt{2} \theta_-, 
\end{eqnarray*}
 
We now introduce $\varphi=-\sqrt{2} \theta_-$ and $n=-\partial_x\phi_-/\sqrt{2}$. In order to discretize, we replace $\partial_x \varphi$ with $(\varphi_{j+1}-\varphi_j)/a$ and $n(ja)$ with $n_j/a$. We end up with
\begin{align*}
  H_-=&\sum_j \left[\frac{u_* K_*}{4\pi a} (\varphi_{j+1}-\varphi_j)^2 +\frac{\pi u_*}{a K_*} n_j^2 \right.\\
  &\qquad\quad- 2 J(t) \rho_0 a \cos \varphi_j\Big],  
\end{align*}
leading to Eq.~(\ref{eq:HLat}) with $c=u_*$ and $K=K_*/2$.

\section{Alternative formulation of SCTDHA for the Frenkel-Kontorova model}\label{app:fk-scha} 

In this appendix we present an alternative formulation of the time dependent self-consistent harmonic approximation for the Frenkel-Kontorova model, following the approach described in Refs.~\cite{nieuwkerk_self-consistent_2019,van_nieuwkerk_low-energy_2020}. As shown in App.~\ref{app:tdva-scha}, this approximation is equivalent to the time-dependent variational principle described in Sec.~\ref{subsec:scha}.  

 This approach relies on approximating the cosine term of the Frenkel-Kontorova model (\ref{eq:HLat}) as
\begin{eqnarray*}
  \cos \varphi_j \simeq e^{-\frac{ \langle \varphi_j^2 \rangle} 2 } \left(1 - \frac {\varphi_j^2} 2 \right), 
\end{eqnarray*}
such that 
\begin{eqnarray}
  \label{eq:scha-ham}
  H_{Lat}\simeq\frac{ c}{2}\sum_{j=1}^{N_s} \left[\frac{\pi n_j^2}{Ka} +\frac {K} {\pi a} (\varphi_{j+1}-\varphi_j)^2 + \rho_0 J(t) a  Z(t) \varphi_j^2  \right], \nonumber \\
\end{eqnarray}
with $Z(t)=e^{-\langle \varphi_j^2 \rangle(t)/2}$. 

It is convenient to use the  Fourier decomposition
\begin{eqnarray*}
  \label{eq:fourier}
  n_j&=&\frac 1 {\sqrt{N_s}} \sum_{k} n_k e^{ikj}, \\
   \varphi_j&=&\frac 1 {\sqrt{N_s}} \sum_{k} \varphi_k e^{ikj}, 
\end{eqnarray*}
where $k=\frac{2\pi p}{N_s}$ with $p$ integer, and to rewrite Eq.(\ref{eq:scha-ham}) as
\begin{eqnarray*}
  \label{eq:scha-fourier}
  H_{Lat}&=&\frac{ c}{2}\sum_k \left\{\frac{\pi n_k n_{-k}}{Ka} \right. \nonumber \\
          &&\left.+ \left[\frac{2K}{\pi a} (1-\cos k) + \rho_0 J(t) a Z(t)\right]  \varphi_k \varphi_{-k} \right\}\nonumber \\. 
\end{eqnarray*}
We obtain the equation of motion
\begin{eqnarray}
  \label{eq:mathieu-like}
  \frac{d^2 \varphi_k(t)}{dt^2} + [\omega^2_k + \Omega^2(t)] \varphi_k(t) =0,    
\end{eqnarray}
with
\begin{eqnarray*}
  \label{eq:omegas-def}
  \omega^2_k=\frac{2c^2}{a^2} (1-\cos ka),\nonumber \\
  \Omega^2(t)=\frac{2\pi c \rho_0 J(t) Z(t)}{K}.  
\end{eqnarray*}
The time evolution of $\varphi_k(t)$ is fully determined by  Eq.~(\ref{eq:mathieu-like}) and the knowledge of $\varphi_k(0)$ and 
\begin{eqnarray}
  \left(\frac{d\varphi_k(t)}{dt}\right)_{t=0} = \frac{\pi c n_k(0)}{Ka}.   
\end{eqnarray}
Those initial conditions are determined by the ground state of the initial Hamiltonian. Since  we have set $\Delta(t<0)=\Delta(0)$, 
\begin{eqnarray*}
  \label{eq:initial-ham}
  H_{SCTDHA}(t<0)=\sum_k \sqrt{\omega^2 + \Omega^2(0)} (a^\dagger_k a_k +1/2), 
\end{eqnarray*}
with
\begin{eqnarray*}
  \varphi_k&=&\frac{a^\dagger_{-k} + a_k}{\sqrt{\frac{2K a}{\pi c}\sqrt{\omega^2_k+\Omega^2(0)}}}, \\
  n_k &=&i \sqrt{\frac{Ka}{\pi c}\sqrt{\omega^2_k+\Omega^2(0)}}\,\frac{a^\dagger_{-k} - a_k}{\sqrt{2}} .
\end{eqnarray*}

Using these expressions and requiring that $Z(t)$ should be determined self-consistently from Eq.~\eqref{eq:scha-ham}, we obtain a set of differential equations for each $k$,
\begin{equation}\label{eq:scha-ode} 
  \frac{d^2 Y_k}{dt^2} +\left[\omega^2_k+ \frac{2\rho_0 \pi c J(t) Z(t) } K \right] Y_k=0,   
\end{equation}
with initial condition
\begin{eqnarray}
  \label{eq:initial_y}
  Y_k(0)&=&\frac{1}{(\omega^2_k+\Omega^2(0))^{1/4}}, \nonumber \\
  \dot{Y}_k(0)&=&-i (\omega^2_k+\Omega^2(0))^{1/4}.
\end{eqnarray}
Finally, the coupling renormalization $Z(t)$ is determined by
\begin{eqnarray}\label{eq:z-y} 
  Z(t)=\exp\left[-\frac{\pi c}{4 N_s K a} \sum_k |Y_k(t)|^2 \right]. 
\end{eqnarray}
Therefore, the equation giving $Z(t)$ is reduced to a nonlinear differential equation for the vector $Y_k(t)$ with initial conditions given by (\ref{eq:initial_y}). Such equation can be integrated numerically with the Runge-Kutta algorithm (see \cite{abramowitz_math_functions} p. 897). The numerical solution of $Y_k(t)$ will be discussed in more detail below.

To fully determine the initial condition, we have to solve for $\Omega(0)$. The selfconsistent equation is
\begin{eqnarray}
  \label{eq:equilibrium-w0}
  &&\frac{a^2\Omega^2(0)}{c^2}=\frac{2\pi \rho_0 J(0)a^2}{cK} \times\nonumber \\
  &&\exp\left[-\frac{\pi}{4N_sK} \sum_k \frac{1}{\sqrt{4\sin^2(ka/2)+(a\Omega/c)^2(0)}}\right].  \nonumber \\
\end{eqnarray}

When $K>\frac 1 8 $, Eq.~(\ref{eq:equilibrium-w0}), has solutions for $J(0)\ll c/(\rho_0 a^2)$ with
\begin{eqnarray*}
  \Omega(0) &=&\frac{2\pi c}a \left(\frac{\rho_0 J(0) a^2}{2\pi c K}\right)^{\frac{4K}{8K-1}}, \\
  Z(0) &=&\left(\frac{\rho_0 J(0) a^2}{2\pi c K}\right)^{\frac 1 {8K-1}} 
\end{eqnarray*}
 If we assume that $Z(t)$ is almost constant, 
the conditions to find resonant modes will be 
\begin{eqnarray*}
  \omega^2(k_*)+\int_0^T \frac{dt}{T} \frac{2\pi \rho_0 c Z(0)}{K} J(t) = \frac{n^2 \gamma^2}{4},  
\end{eqnarray*}
with $n\ge 1$ integer,  so for
\begin{eqnarray}\label{eq:stab-cond} 
  \gamma > 2 \sqrt{\omega^2(\pi)+\int_0^T \frac{dt}{T} \frac{2\pi \rho_0 c Z(0)}{K} J(t)},  
\end{eqnarray}
resonant modes are absent. In such case of high frequency
driving, $Z(t)$ can reach a non-zero limit at long times.  However,
when the condition~(\ref{eq:stab-cond})is not satisfied, an unstable
mode growing exponentially will be present, leading to $Z(t) \to
0$. Thus, the approximation of constant $Z(t)$  breaks down over a
timescale determined by the Floquet exponent with the largest
imaginary part. Besides, the previous argument also shows that at
long times, when $\gamma<2\omega(\pi)$, $Z(t)$ cannot have a strictly
positive limit. Otherwise, we would find an exponentially growing mode
leading to $Z(t) \to 0$ and a contradiction.  Thus, we are lead to
expect two regimes, one of high frequency driving with $Z(t)$
approaching a finite limit at long times, and another of low frequency
driving with $Z(t)$ going to zero for long times. In
App.~\ref{app:asymp-Z}, we consider a simplified model, where only the
resonant modes are taken into account. The model shows some periodic
revivals of the coherence; albeit with a duration that decreases as
$O(1/t)$.\\

\subsection{Numerical results for renormalization term $Z(t)$}

\begin{figure}[b]
    \centering
    \includegraphics[width=\columnwidth]{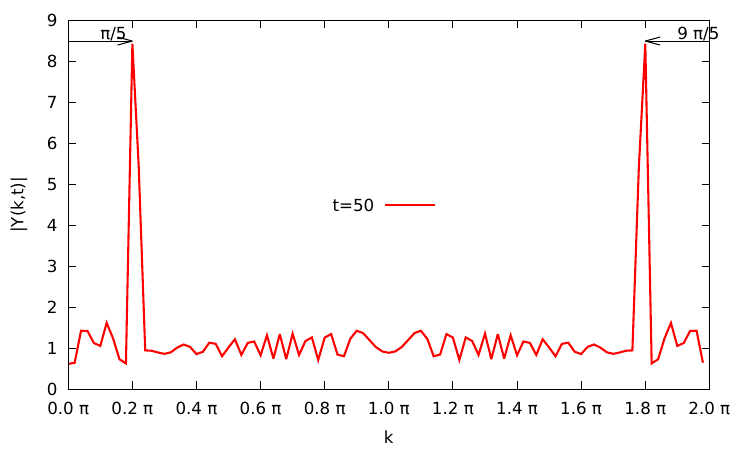} 
    \caption{Plot of the modulus of $Y_k(t)$ versus $k$ at time $t=50$. The two resonant modes at $k^*=\frac \pi 5$ and $2\pi -k^*$ are clearly visible.}
    \label{fig:modes-final} 
  \end{figure}

  Assuming for a moment that $Z(t)$ is given, we consider the linear differential equation
\begin{eqnarray*}\label{eq:linear-ode} 
   \frac{d^2 Y_k}{dt^2} + [\omega^2_k + \Omega^2(t)] Y_k(t) =0,
\end{eqnarray*}
and its fundamental solutions
\begin{equation*}
  \label{eq:fundamental}
  \left(
    \begin{array}{cc}
      y_1 & y_2 \\
      \dot{y}_1 & \dot{y}_2 
    \end{array}
\right)(k,t=0) =  \left(
    \begin{array}{cc}
      1 & 0 \\
      0 & 1 
    \end{array}
\right).
\end{equation*}
We find that
\begin{eqnarray*}
 \varphi_k(t)&=&\frac{1}{(2\frac{K a}{\pi c} \sqrt{\omega^2_k +\Omega^2(0)})^{1/2}} \left[a_k (y_1-i  \sqrt{\omega^2_k +\Omega^2(0)} y_2) \nonumber \right. \\ &&\left.+ a^\dagger_{-k}   (y_1 +i  \sqrt{\omega^2_k +\Omega^2(0)} y_2)\right],  
\end{eqnarray*}
 resulting in
\begin{eqnarray*}\label{eq:scha-Z} 
  Z(t)=\exp\left[-\frac{\pi c}{4N K a} \sum_k \frac{y_{1k}(t)^2 + y_{2k}(t)^2 (\omega^2_k + \Omega^2(0))}{\sqrt{\omega^2_k+\Omega^2(0)}}\right].  
\end{eqnarray*}

We have solved numerically the differential equation for $Z(t)$ and $Y_k(t)$ for $J(t)=J_0+J_1\cos(\gamma t)$, where $\gamma T=2\pi$. We have picked our unit of time so that $c/a=1$ and set $K=\pi$. We show the results for the case of $J_0=J_1=0.1$ and $\gamma=1.5$. In figure \ref{fig:modes-final}, the modulus of $Y_k(t)$ is plotted as a function of $k$. We have resonant modes at $k^*=\pi/5$ and $2\pi-k^*=9\pi/5$.   

\begin{figure}[t]
    \centering
    \includegraphics[width=\columnwidth]{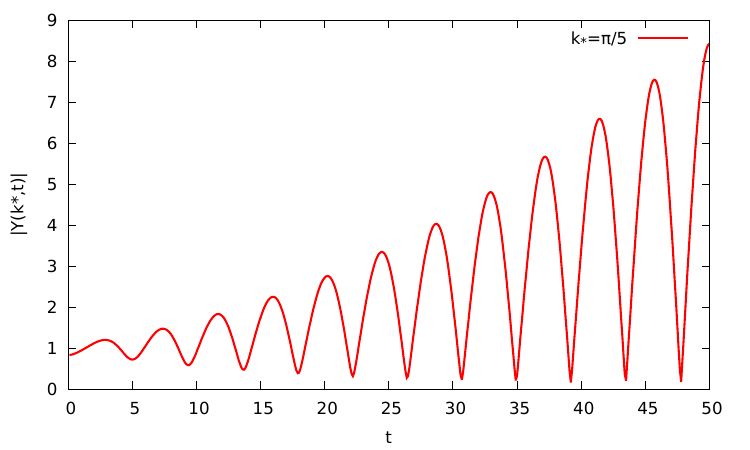} 
    \caption{Plot of $|Y(k^*,t)|$ as a function of time. The amplitude of the oscillations is increasing with time. }
    \label{fig:resonant}
  \end{figure}

  The time evolution of the resonant mode is shown on Fig.~\ref{fig:resonant}. The amplitude is not a monotonous function of time, but shows a succession of maxima and minima, with the height of the maxima increasing with time.  
  
\begin{figure}[b]
    \centering
    \includegraphics[width=\columnwidth]{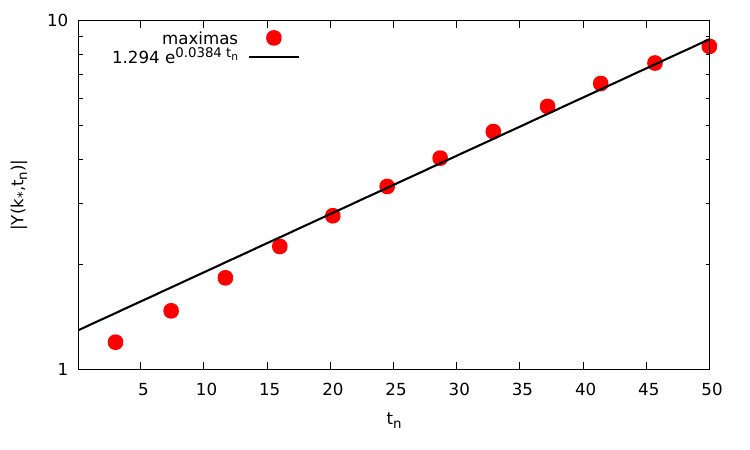} 
    \caption{Plot of $|Y_{k^*}(t_{max,n})|$ versus $t_{max,n}$ on semi-logarithmic scale. This can be fitted by an exponential.}
    \label{fig:maximas}
  \end{figure}

 Plotting the maxima as a function of time on a semi-logarithmic scale (Fig.~\ref{fig:maximas}), we find the points are close to a straight line, compatible with exponential growth at short times.
 
\subsection{Analytical results for the self-consistent time-dependent harmonic approximation}\label{app:asymp-Z} 

We now discuss a few analytical approximations derived within the SCTDHA. For the differential equation~\eqref{eq:scha-ode} we can show that for each $k$,
\begin{equation*}
  Y^*_k(t) \frac{dY_k(t)}{dt} - Y_k(t) \frac{dY_k(t)^*}{dt} = -2i,  
\end{equation*}
and we have energy conservation
\begin{eqnarray}\label{eq:energy-conservation} 
&&  \frac d {dt} \left(\sum_k \left|\frac{dY}{dt}\right|^2(k,t) + \omega^2_k |Y_k(t)|^2 -8 \Delta(t) Z(t)\right) \nonumber \\ && = -8 N \frac{d\Delta}{dt} Z(t),   
\end{eqnarray}
where $\Delta(t)=\rho_0 a J(t)$.  
After integrating~(\ref{eq:energy-conservation}) over time, we find that if there is at least one  mode $Y(k\ne 0,t)$ whose amplitude grows to infinity, then
\begin{eqnarray*}
  \int_0^{+\infty} Z(t) \frac{d\Delta}{dt} dt = -\infty, 
\end{eqnarray*}
and since $0<Z(t)<1$ while $\Delta(t)$ is also bounded, 
\begin{eqnarray*}
  \int_0^{+\infty} \Delta(t) \frac{dZ}{dt} dt = +\infty.  
\end{eqnarray*}
Using 
\begin{eqnarray*}
  \left|\int_0^{t} Z(t') \frac{\Delta}{dt'}\right|<\int_0^{t} Z(t) \left|\frac{\Delta}{dt}\right| <\\\qquad\qquad < \mathrm{Max}_{0<t<T}\left|\frac{d\Delta}{dt}\right| \int_0^{t} Z(t') dt',   \nonumber 
\end{eqnarray*}
so if we have modes with a divergent amplitudes, 
\begin{equation*}
\int_0^{+\infty} Z(t) dt = +\infty.     
\end{equation*}
Eq.~(\ref{eq:z-y}), however, implies that having modes with divergent amplitudes leads to $Z(t)\to 0$. 
So $Z(t)$ should decay to zero sufficiently slowly at long times to yield a divergent integral, implying that the amplitudes of unstable modes $|Y(k,t)|$ are also growing sufficiently slowly. 
Now, let's return to Eq.~(\ref{eq:scha-ode}),  with $J(t)$ given by Eq.~(\ref{eq:Jt}), and let's consider $k=k_r$ where $\omega(k_r)=\gamma/2$. We will seek $Y(k_r,t)$ in the form
\begin{eqnarray*}
  Y(k_r,t)=A(t) e^{i\frac\gamma 2 t} + B(t) e^{-i\frac \gamma 2 t},  
\end{eqnarray*}
neglecting $2\pi\rho_0c J_0 Z(t)/K$ compared with $\gamma^2/4$. We end up with the system of differential equations,
\begin{eqnarray*}
  &&  \frac{d^2 A}{dt^2} + i\gamma \frac{dA}{dt} -i \frac{\pi \rho_0 c  J_1} K Z(t) B =0, \\
   &&  \frac{d^2 B}{dt^2} - i\gamma \frac{dB}{dt} + i \frac{\pi \rho_0 c  J_1} K Z(t) A =0.
\end{eqnarray*}
Neglecting the second derivatives, as $A(t)$ and $B(t)$ are expected to vary slowly over one period, and introducing Pauli matrices, we rewrite our system as
\begin{equation*}
  i\gamma \sigma_3 \left(\begin{array}{c} A \\ B \end{array}\right) +\frac{\pi \rho _0 c J_1}{K} Z(t) \sigma_2  \left(\begin{array}{c} A \\ B \end{array}\right)=0, 
\end{equation*}
that is solved in the form
\begin{equation*}\label{eq:matrix-exponential} 
   \left(\begin{array}{c} A(t) \\ B(t) \end{array}\right)=\exp\left[\int_{t_0}^t \frac{\pi \rho_0c J_1}{\gamma K} \sigma_1 Z(t') dt'\right]  \left(\begin{array}{c} A(t_0) \\ B(t_0) \end{array}\right). 
\end{equation*}
We see that when the integral of $Z$ is finite, $A$ and $B$ remain finite, but if the integral of $Z$ is divergent, $A$ and $B$ will grow to infinity.
With the help of (\ref{eq:matrix-exponential}), we find
\begin{equation*}
  Y(k_r,t)=\mathcal{A} \cos \left(\frac{\gamma t} 2\right) \exp\left[\int_0^t\frac{\pi \rho_0 c J_1}{\gamma K} Z(t') dt'\right] +\ldots  
\end{equation*}
The same equation holds for $Y(-k_r,t)$. The amplitude squared diverges as
\begin{equation*}
   |Y(k_r,t)|^2=\frac{|\mathcal{A}|^2}2 [1 - \cos \gamma t]  \exp\left[\int_0^t\frac{2\pi \rho_0 c  J_1}{\gamma K} Z(t') dt'\right] +\ldots ,  
 \end{equation*}
 and we have 
 \begin{align}\label{eq:kin_en-Z} 
   \left|\frac{dY(k_r,t)}{dt}\right|^2 & +\frac {\gamma^2} 4 |Y(k_r,t)|^2 =\nonumber\\
   &\frac{\gamma^2 |\mathcal{A}|^2}{4} \exp\left[\int^t \frac{2\pi \rho_0 c J_1}{\gamma K} Z(t') dt'\right].   
 \end{align}
 Assuming only the modes at $\pm k_r$ are divergent, and introducing
 \begin{equation}
   \label{eq:zeta-def}
   \zeta(t) =\frac{\pi c |\mathcal{A}|^2}{4N K a}  \exp\left[\int_0^t\frac{2\pi \rho_0 c J_1}{\gamma K} Z(t') dt'\right], 
 \end{equation}
 we obtain
 \begin{equation*}\label{eq:zeta-ode} 
   \frac{\dot{\zeta}}{\zeta} = \frac{2\pi \rho_0 c  J_1}{\gamma K } \exp\left[-(1-\cos \gamma t) \zeta(t) \right] .
 \end{equation*}
 We have the inequality 
 \begin{equation*}
   \int^{\zeta(t)} \frac{d\zeta'}{\zeta'} e^{2\zeta'} \ge \frac{2\pi J_1 \rho_0 c t}{\gamma K},  
 \end{equation*}
 yielding
 \begin{equation*}
   \zeta(t)\ge \frac 1 2 \ln \left[\frac{2\pi  J_1 \rho_0 c t}{\gamma K} \ln\left(\frac{2\pi J_1 \rho_0 c  t}{\gamma K}\right)\right]+o(1).  
 \end{equation*}
 Since $\zeta(t) \to +\infty$,  Eq.~(\ref{eq:zeta-ode}) implies that except in the vicinity of $\gamma t = 2 n \pi$, with integer $n$, $\dot{\zeta}\ll 1$. Near $\gamma t =2 n \pi$, we can expand the cosine in Eq.~(\ref{eq:zeta-ode}) to obtain
 \begin{equation*}
   \frac{\dot{\zeta}}{\zeta} \simeq e^{-\zeta(2n\pi/\gamma) (\gamma t -2 n \pi)^2/2}.  
   \end{equation*}
 
 By integrating, we have the relations 
   \begin{eqnarray*}
     \zeta\left(\frac{2n+1}{\gamma}\right) &\simeq& \zeta\left(\frac{2n}{\gamma}\right) \exp\left[\frac{\pi \rho_0 c J_1}{\gamma^2 K}\sqrt{\frac{2\pi}{\zeta\left(\frac{2n}{\gamma}\right)}} \right],\nonumber \\
       \zeta\left(\frac{2n-1}{\gamma}\right) &\simeq& \zeta\left(\frac{2n}{\gamma}\right) \exp\left[-\frac{\pi \rho_0 c J_1}{\gamma^2 K}\sqrt{\frac{2\pi}{\zeta\left(\frac{2n}{\gamma}\right)}} \right], 
   \end{eqnarray*}
   provided $\zeta(2n\pi) \gg 1$. Using the above expressions to obtain $\zeta[(2n+1)\pi/\gamma$ as a function of $\zeta[2n\pi/\gamma]$ and of $\zeta[(2n+2)\pi/\gamma]$, we get
   \begin{widetext}
   \begin{eqnarray*}
  \zeta\left(\frac{2n}{\gamma}\right) \exp\left[\frac{\pi \rho_0 c J_1}{\gamma^2 K}\sqrt{\frac{2\pi}{\zeta\left(\frac{2n}{\gamma}\right)}} \right]=    \zeta\left(\frac{2n+2}{\gamma}\right) \exp\left[-\frac{\pi \rho_0 c J_1}{\gamma^2 K}\sqrt{\frac{2\pi}{\zeta\left(\frac{2n+2}{\gamma}\right)}} \right].
   \end{eqnarray*}
 \end{widetext}
 We can rewrite this equation in differential form to arrive at
   \begin{eqnarray*}
     \zeta\left(\frac{2n}{\gamma}\right) \simeq 2\pi \left(\frac{n\pi \rho_0 c J_1}{\gamma^2 K}\right)^2,     
   \end{eqnarray*}
   and for $(2n-1)\pi<\gamma t<(2n+1)\pi$, 
   \begin{align*}
     \zeta(t) \simeq & 2\pi \left(\frac{n\pi \rho_0 c J_1}{\gamma^2 K}\right)^2\times \\
     &\exp\left\{\frac{1} {2n} \mathrm{erf}\left[\frac{2\pi^{3/2} \rho_0 c J_1}{\gamma K} n\left(t-\frac{2n \pi}{\gamma}\right)\right]\right\},\nonumber \\ 
   \end{align*}
   and
   \begin{equation*}
     Z(t) \simeq \exp\left[-\pi \left(\frac{\pi \rho_0 c J_1 n}{\gamma K}\right)^2 \left(t - 2n \frac \pi \gamma\right)^2\right], .
   \end{equation*}
   We thus obtain
   \begin{equation*}
     \int_{\frac{(2n-1)\pi}{\gamma}}^{\frac{(2n+1)\pi}{\gamma}}dt Z(t) \simeq \frac{\gamma K}{\pi \rho_0 c J_1 n},   
   \end{equation*}
   and
   \begin{equation*}
     \int_0^t dt Z(t) \sim \frac{\gamma K}{\pi \rho_0 c J_1} \ln \frac{\gamma t}{2\pi}.  \end{equation*}
   Qualitatively, at long times, $Z(t)$ is nearly zero except in intervals of size $\sim 1/n$ around $t=2n\pi/\gamma$, where $Z(t)\simeq 1$. This leads to short lived periodic revivals of the phase coherence between the chains. The non-resonant modes lead to a blurring of those revivals, but spikes in $Z(t)$ are visible in our numerical simulations.  
   If we consider the total energy, $\Delta(t) Z(t)$ is bounded, so it is enough to use~(\ref{eq:kin_en-Z}) to find the energy going as
   \begin{align*}
   \langle H(t)\rangle \sim & \frac{2N a}{cK} \left(\frac{\pi \rho_0 c J_1 n}{\gamma}\right)^2 \times\\
   &\exp\left\{\frac 1{2n} \mathrm{erf}\left[\frac{2\pi^{3/2} \rho_0 c J_1}{\gamma K} n \left(t-\frac {2 n \pi}\gamma\right) \right]\right\},  
\end{align*}
when $t\sim 2n\pi$. We find an energy increasing as $O(t^2)$.

\section{Equivalence of the \li{time-dependent Gaussian variational ansatz} and the self-consistent time dependent harmonic approximation}\label{app:tdva-scha}

In this appendix we show the equivalence of the time-dependent variational principle outlined in Sec.~\ref{subsec:scha}, and the time dependent self-consistent harmonic approximation as formulated in Refs.\cite{nieuwkerk_self-consistent_2019,van_nieuwkerk_low-energy_2020} and used in App.~\ref{app:fk-scha}.

First, expressing $\Sigma_k$ as a function of $\dot{G}_k/G_k$,  the system (\ref{eq:EOM}) is rewritten as a single second order differential equation for $G_k$
\begin{eqnarray}
  \label{eq:tdva-2nd_order}
    \frac K {4\pi c} \left[\frac {\ddot{G}_k}{G_k} -\frac 1 2 \left(\frac{\dot{G}_k}{G_k}\right)^2\right] =\frac {\pi c} {8K G_k^2} -\frac {cK} {2\pi} k^2 - J(t) Z(t)\rho_0.\nonumber \\
\end{eqnarray}
Introducing $y_k=\sqrt{G_k}$, one has   
\begin{eqnarray*}
  \frac 1 {y_k} \frac{d^2 y_k}{dt^2} =\frac 1 2 \left[\frac {\ddot{G}_k}{G_k} -\frac 1 2 \left(\frac{\dot{G}_k}{G_k}\right)^2\right],
\end{eqnarray*}
so that in terms of $y_k$, Eq.~(\ref{eq:tdva-2nd_order}) becomes
\begin{equation}
  \label{eq:tdva-y}
   \frac{d^2 y_k}{dt^2} = \frac {(\pi c)^2} {4K^2 y_k^3} -\left(c^2 k^2 + \frac{2\pi c J(t) Z(t) \rho_0}{K}\right) y_k.
\end{equation}
Eq.~(\ref{eq:tdva-y}) is the equation of motion of a classical particle moving in a time-dependent harmonic central force field written in polar coordinates\cite{landau1976mechanics}. The term $\pi^2 c^2/(4K^2 y_k^3)$ is the centrifugal force, while the term proportional to $y_k$ is the harmonic restoring force. Eq.~(\ref{eq:tdva-y}) is thus linearized by introducing the angular coordinate $\theta_k$ satisfying
\begin{equation}
\frac{d \theta_k}{dt}=\frac{\pi c}{2K y_k^2}.
\end{equation}
 with initial condition $\theta_k(0)=0$. In terms of the variable $z_k=\sqrt{G_k}e^{i\theta_k}$, Eq..~(\ref{eq:tdva-y}) becomes 
\begin{equation}
  \label{eq:tdva-linearized}
  \frac{d^2 z_k}{dt^2} = \left(c^2 k^2 +\frac{2\pi c J(t) Z(t)\rho_0} K\right)  z_k. 
\end{equation}
That is precisely Eq.~\eqref{eq:scha-ode}, obtained using the selfconsistent time dependent harmonic approximation (SCTDHA) \cite{nieuwkerk_self-consistent_2019,van_nieuwkerk_low-energy_2020}.

Having reduced Eq.~(\ref{eq:EOM}) to a linear second order equation, the initial conditions of Eq.~(\ref{eq:tdva-linearized}) are
\begin{eqnarray*}
  \label{eq:initial-tdva}
  z_k(0)&=&\sqrt{G_k(0)} \\
 \dot{z}_k(0)&=&\frac {2\pi c} K \sqrt{G_k(0)} \Sigma_k(0) + \frac {i\pi c} {2K \sqrt{G_k(0)}}.
\end{eqnarray*}
Again using the solutions $y_{1k}(t)$ and $y_{2k}(t)$ of the differential equation
\begin{eqnarray*}
  \frac{d^2 Y_k}{dt^2} = -\left(c^2 k^2 +\frac{2\pi c J(t) Z(t)\rho_0} K\right) Y_k,
\end{eqnarray*}
with initial conditions
\begin{equation*}
\left(\begin{array}{cc} y_1(0) & y_2(0) \\ \dot{y}_1(0) & \dot{y}_2(0) \end{array} \right) = \left(\begin{array}{cc} 1 & 0 \\ 0 & 1 \end{array}\right),
\end{equation*}
we find
\begin{eqnarray*}
 z_k(t)=\sqrt{G_k(0)} \left[y_{1k}(t) + \frac {2\pi c} K \Sigma_k(0) y_{2k}(t) + \frac {i\pi c y_{2k}(t)}{2KG_k(0)}  \right]. \nonumber \\
\end{eqnarray*}
Since our initial state was a steady state, $\Sigma_k(0)=0$. Then, $G_k(t)=|z_k(t)|^2$ yields
\begin{eqnarray*}
 G_k(t)=G_k(0) \left[ y_{1k}(t)^2 + \frac{\pi^2 c^2 y_{2k}(t)^2}{4K^2 G_k(0)^2} \right],
\end{eqnarray*}
and using the equilibrium Green's function Eq.~(\ref{eq:tdva_green-eq}),  
we get
\begin{eqnarray*}\label{eq:z-tdva}
 && Z(t)= \nonumber \\
  && \exp\left[-\int_{-\Lambda}^\Lambda \frac{dk}{8  K \sqrt{k^2 +\Delta_0^2}} \left(y_{1k}(t)^2 + c^2 (k^2+\Delta_0^2) y_{2k}(t)^2\right) \right].\nonumber \\
\end{eqnarray*}

\section{Gap renormalization and infinite temperature runaway instability in SCTDHA in the amplitude-modulated sine-Gordon model}\label{app:gapRenorm}

In Sec.~\ref{sec:gapped} we approximated the gap of the sine-Gordon model with  $J_0\neq 0$ by Eq.~\eqref{eq:egap}. A more precise expression within SCTDHA takes into account the renormalization of the tunnel coupling by quantum fluctuations. To this end, the system is assumed to be in its stationary gapped ground state, i.e. $\Sigma_k = 0$ for $t \leq 0$. The initial correlator for the continuum version is determined from Eq.~\eqref{eq:sigdot},
\begin{equation*}
G_k = \frac{\pi}{2K}\,\frac{1}{\sqrt{k^2 + \Delta_0^2}},
\end{equation*}
where $\Delta_0$ is self-consistently given as:
\begin{equation*}
\Delta_0^2 = \frac{2\pi J \rho_0}{c K}\,\exp\left(-\frac{1}{2}\int_{-\Lambda}^\Lambda \frac{\dd k}{4K} \,\frac{1}{\sqrt{k^2 + \Delta_0^2}}\right).
\end{equation*}
Assuming $\Delta_0 \ll \Lambda$, the above equation gives $\Delta_0^2 \approx (g_0/2)[\Delta_0/(2\Lambda)]^{1/(4 K)}$, which implies that $K_c =\frac{1}{8}$ (Kosterlitz-Thouless transition). The cosine is relevant (irrelevant) for $K>K_c$ $(K<K_c)$. In this initial gapped phase, the classical oscillation frequency,
$\sqrt{g_0}$ is renormalized by the factor $Z$ due to quantum fluctuations.

In the presence of a modulation in the tunnel coupling, the system can enter a region of infinite temperature runaway instability depending on the strength and frequency of the modulation. This effect can be understood within  the variational wavefunction approach as follows. In the initial gapped phase, the classical oscillation frequency near the $\varphi=0$ minimum of the cosine potential is renormalized by the factor $Z$ due to quantum fluctuations. The introduction of a modulation to the amplitude of the bare cosine potential can give rise to an ergodic regime which amplifies quantum fluctuations (i.e. leads to ``particle generation'' via parametric resonance) and closes the gap, i.e. $Z(t) \rightarrow 0$. Once the gap closes, it remains closed; we take this as an indication of the runaway to the infinite-temperature limit. For weaker modulations, however, the system can stay in a non-ergodic regime, where the quantum fluctuations remains bounded, $Z(t)$ remains finite at all times, and $\varphi$ remains localized. The phase diagram showing the ergodic-non-ergodic phase transition was previously derived in Ref.~\onlinecite{CITRO2015694}.\\

 \section{Adiabatic limit in the selfconsistent time dependent harmonic approximation}\label{app:adiabatic}

If we make a WKB approximation\cite{landau_mecaq} in Eq.~(\ref{eq:mathieu-like}), we find:
\begin{eqnarray*}
  \label{eq:wkb}
&&  y_{1k}(t)=\left(\frac{\omega_k^2 +\Omega^2(0)}{\omega_k^2 + \Omega^2(t)}\right)^{1/4} \cos \left[ \int_0^t \sqrt{\omega_k^2 +\Omega^2(t')} dt'\right], \nonumber \\ 
&&  y_{2k}(t)=\frac {(\omega_k^2 +\Omega^2(0))^{-1/4}} {(\omega_k^2 + \Omega^2(t))^{1/4}} \sin \left[ \int_0^t \sqrt{\omega_k^2 +\Omega^2(t')} dt'\right],\nonumber \\
\end{eqnarray*}
and substituting this expression  into~(\ref{eq:z-y}) gives
\begin{eqnarray*}
  \Omega(t)=\frac{2\pi c} a \left(\frac{\rho_0 a^2 J(t)}{2\pi c K}\right)^{\frac{4K}{8K-1}}.  
\end{eqnarray*}
In the adiabatic regime, the modes adapt instantaneously to the variations of the hopping $J(t)$ and the evolution of the gap is adiabatic.
Now, when $J(t+T)=J(t)$ is periodic,  $\Omega(t)=\Omega(t+T)$.  The WKB solutions~(\ref{eq:wkb}) can then be combined as $y_1\pm i (\omega_k^2+\Omega^2(0))^{1/2} y_2$  to satisfy a Floquet condition with Floquet exponent
\begin{equation*}
  \pi \nu(k)= \pm \int_0^T \sqrt{\omega_k^2 + \Omega^2(t)} dt.  
\end{equation*}
The condition for the validity of the WKB approximation is
\begin{eqnarray*}
 \left| \frac{d}{dt} (\sqrt{\omega_k^2 +\Omega^2(t)})\right| \ll \omega_k^2 +\Omega^2(t),    
\end{eqnarray*}
so it is always valid for $\omega_k\gg \dot{\Omega}{\Omega}$. For $k\to 0$, the condition  becomes
\begin{eqnarray}\label{eq:wkb-condition} 
  \left|\frac{d\Omega}{dt}\right| \ll  \Omega^2(t).    
\end{eqnarray}
In the case of a periodic $\Omega(t)>0$,  Eq.~(\ref{eq:wkb-condition}) imposes $\gamma < \mathrm{min}_{0<t<T} \Omega(t)$. The resonance condition being $\sqrt{\omega_k^2+\Omega^2} = n\gamma/2$, the only possible resonances are at $n\gg 1$ and their width and Lyapunov exponent are exponentially suppressed with $\Omega/\gamma$. This explains the purely real Floquet exponents in the adiabatic limit. 
Eq.~(\ref{eq:wkb-condition}) also indicates that adiabaticity breaks down when $\Omega(t) \to 0$.  
In the case of $J(t)=\lambda (t_0-t)$, we enter a non-adiabatic regime at $t=t_*$ with 
\begin{eqnarray}
  t_0-t_* \sim \frac a c \left(\frac{c^2 K}{\rho_0 a^3 \lambda}\right)^{\frac{4K}{12K-1}}.  
\end{eqnarray}
In particular the modes with momenta  $k<k_*  \sim (\rho_0 a^3 \lambda/c^2)^{\frac{4K}{12K-1}}$ start to exhibit non-adiabatic evolution. But since the timescale $t_0-t_*$ is less than one period of these modes, they behave as if they had been frozen at time $t=t_*$. As a result, $\langle \cos \varphi_j\rangle(t_0)=Z(t_*)\sim \lambda^{\frac 1 {12K-1}}$ indicating a residual coherence.

\bibliography{modsineGordon}
\end{document}